%% file: main.tex
\documentclass[sigplan,non-acm]{acmart}
\newcommand{\asplossubmissionnumber}{128}
\input{packages}
\input{defs}

\AtBeginDocument{%
}
\copyrightyear{2025} 
\acmYear{2025} 
\setcopyright{rightsretained}
\acmConference[ASPLOS '25]{Proceedings of the 30th ACM International Conference on Architectural Support for Programming Languages and Operating Systems, Volume 1}{March 30--April 3, 2025}{Rotterdam, Netherlands}
\acmBooktitle{Proceedings of the 30th ACM International Conference on Architectural Support for Programming Languages and Operating Systems, Volume 1 (ASPLOS '25), March 30--April 3, 2025, Rotterdam, Netherlands}
\acmDOI{10.1145/3669940.3707226}
\acmISBN{979-8-4007-0698-1/25/03}

\makeatletter
\gdef\@copyrightpermission{
  \begin{minipage}{0.3\columnwidth}
    \href{https://creativecommons.org/licenses/by/4.0/}
    {\includegraphics[width=0.90\textwidth]{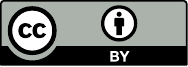}}
  \end{minipage}\hfill
  \begin{minipage}{0.7\columnwidth}
    \href{https://creativecommons.org/licenses/by/4.0/}
    {This work is licensed under a Creative Commons Attribution International 4.0 License.}
  \end{minipage}
  \vspace{5pt}
}
\makeatother

\begin{document}

\newpage

\title[\pname{}: Exploiting Temporal Patterns for All-Resource Oversubscription]{\pname{}: Exploiting Temporal Patterns for All-Resource Oversubscription in Cloud Platforms}

\author{\vspace{15pt}Benjamin Reidys}
\affiliation{%
    \country{University of Illinois Urbana-Champaign, USA}
}
%\author{Benjamin Reidys$^{\dagger}$\FootnotemarkAAffil}
\authornote{
%Benjamin Reidys, Haoran Ma, Stanko Novaković, and Lisa Hsu are affiliated with UIUC, UCLA, Google, and Meta respectively.
%Reidys, Ma, Novaković, and Hsu are affiliated with UIUC, UCLA, Google, and Meta respectively.
%They were all at Microsoft during this work.
%Ma, Novaković, and Hsu are affiliated with UCLA, Google, and Meta respectively.
Benjamin Reidys and Haoran Ma interned at Microsoft.
Stanko Novaković and Lisa Hsu were at Microsoft when they contributed to this work.
%\vspace{-5pt}
}
\author{\vspace{15pt}Pantea Zardoshti}
\affiliation{%
    \country{Microsoft\\ Redmond, USA}
}
\author{\vspace{15pt}\'I\~nigo Goiri}
\affiliation{%
    \country{Microsoft\\ Redmond, USA}
}
\author{\vspace{15pt}Celine Irvene}
\affiliation{%
    \country{Microsoft\\ Redmond, USA}
    \vspace{5pt}
}

\author{Daniel S. Berger}
\affiliation{%
    \country{Microsoft, Redmond, USA}
    \country{University of Washington, Seattle, USA}
    %\country{Microsoft\\ Redmond, USA\\}
    %\country{University of Washington\\ Seattle, USA}
}
\author{Haoran Ma$^{*}$}
\affiliation{%
    \country{University of California-Los Angeles, USA}
}
\author{Kapil Arya}
\affiliation{%
    \country{Microsoft\\ Redmond, USA}
}
\author{Eli Cortez}
\affiliation{%
    \country{Microsoft\\ Redmond, USA}
    \vspace{5pt}
}

\author{Taylor Stark}
\affiliation{%
    \country{Microsoft\\ Redmond, USA}
}
\author{Eugene Bak}
\affiliation{%
    \country{Microsoft\\ Redmond, USA}
}
\author{Mehmet Iyigun}
\affiliation{%
    \country{Microsoft\\ Redmond, USA}
}
\author{Stanko Novaković$^{*}$}
\affiliation{%
    \country{Google\\ Mountain View, USA}
    \vspace{5pt}
}

\author{Lisa Hsu$^{*}$}
\affiliation{%
    \country{Meta\\ Menlo Park, USA}
}
\author{Karel Trueba}
\affiliation{%
    \country{Microsoft\\ Redmond, USA}
}
\author{Abhisek Pan}
\affiliation{%
    \country{Microsoft\\ Redmond, USA}
}
\author{Chetan Bansal}
\affiliation{%
    \country{Microsoft\\ Redmond, USA}
    \vspace{5pt}
}

\author{Saravan Rajmohan}
\affiliation{%
    \country{Microsoft\\ Redmond, USA}
}
\author{Jian Huang}
\affiliation{%
    \country{University of Illinois Urbana-Champaign, USA}
}
\author{Ricardo Bianchini}
\affiliation{%
    \country{Microsoft\\ Redmond, USA}
    % Adding some breathing space between authors and text
    \vspace{15pt}
}

\date{}


\renewcommand{\shortauthors}{Benjamin Reidys et al.}

\input{abstract}

\maketitle

\pagenumbering{gobble}

%\pagestyle{empty}

\input{intro}
\input{characterization}

\input{design}

\input{lessons}

\input{impl}
\input{eval}
\input{related}

\input{conclusion}
\input{acks}

\bibliographystyle{plain}
\balance
\bibliography{references}

\end{document}

%% file: packages.tex
\usepackage{hyperref}
\usepackage[normalem]{ulem}
\usepackage{xcolor}
\usepackage{amsmath,amsfonts}
\usepackage{algorithmic}
\usepackage{graphicx}
\usepackage{textcomp}
\usepackage{enumitem}
\usepackage{listings}
\usepackage{subcaption}
\usepackage{soul}
\usepackage{comment}
\usepackage{tikz}
\usepackage{xspace}
\usepackage{array}
\hypersetup{
	unicode=true,
        bookmarksnumbered=true,
	colorlinks=true,
	linkcolor={red!70!black},
	citecolor={red!70!black},
	urlcolor={blue!70!black},
	pdfborder={0 0 0}
}
\usepackage[capitalize,nameinlink]{cleveref}
\usepackage{flushend}
\usepackage{pifont}

%% file: defs.tex
\newcommand{\myparagraph}[1]{\vspace{\smallskipamount}\noindent\textbf{#1.\xspace}}
\newcommand{\myparagraphnodot}[1]{\vspace{\smallskipamount}\noindent\textbf{#1\xspace}}
\newcommand{\myparagraphemph}[1]{\vspace{\smallskipamount}\noindent\emph{#1.\xspace}}
\newcommand{\eg}{\emph{e.g.}\xspace}

\newcommand{\ie}{\emph{i.e.}\xspace}

\newcommand*{\rom}[1]{\uppercase\expandafter{\romannumeral #1\relax}}
\newcommand{\insightparagraph}[1]{\vspace{\smallskipamount}\noindent\textbf{\emph{Insight \rom{#1}:\xspace}}}

\newif\ifshowcomment
\showcommenttrue
\ifshowcomment
    \newcommand{\ben}[1]{{\color{orange}[Ben: #1]}}
    \newcommand{\inigo}[1]{{\color{blue}[IG: #1]}}
    \newcommand{\pantea}[1]{{\color{teal}[PZ: #1]}}
    \newcommand{\stanko}[1]{{\color{teal}[SN: #1]}}
    \newcommand{\ricardo}[1]{{\color{cyan}[RB: #1]}}
    \newcommand{\jian}[1]{{\color{violet}[Jian: #1]}}
    \newcommand{\celine}[1]{{\color{green}[Celine: #1]}}
    \newcommand{\todo}[1]{{\color{red}[TODO: #1]}}
\else
    \newcommand{\ben}[1]{}
    \newcommand{\inigo}[1]{}
    \newcommand{\pantea}[1]{}
    \newcommand{\stanko}[1]{}
    \newcommand{\ricardo}[1]{}
    \newcommand{\jian}[1]{}
    \newcommand{\celine}[1]{}
    \newcommand{\todo}[1]{}
\fi

\newcommand{\pname}[1]{{{{Coach}}}{#1}}
\newcommand{\implvm}[1]{{{{CoachVM}}}{#1}}
\newcommand{\cimplvm}[1]{{{{CVM}}}{#1}}
\newcommand{\cimplvml}[1]{{{{Cvm}}}{#1}}
\newcommand{\gpvm}[1]{{{{Gpvm}}}{#1}}

\newcommand{\goalmincust}[1]{{{{G1}}}{#1}}
\newcommand{\goalminworkl}[1]{{{{G2}}}{#1}}
\newcommand{\ovm}[1]{{{{Ovm}}}{#1}}

\newcommand{\cmark}{\ding{51}}
\newcommand{\xmark}{\ding{55}}
\newcommand{\greencmark}{\color{green!45!black}{\cmark}}
\newcommand{\redxmark}{\color{red!80!black}{\xmark}}
\newcommand{\customtimes}{\,\tikz \draw (0,0) -- (0.15,0.15) (0,0.15) -- (0.15,0);\,}

%% file: abstract.tex
\begin{abstract}
Cloud platforms remain underutilized despite multiple proposals to improve their utilization (\eg{}, disaggregation, harvesting, and oversubscription). 
Our characterization of the resource utilization of virtual machines (VMs) in Azure reveals that, while CPU is the main underutilized resource, we need to provide a solution to manage all resources holistically.
We also observe that many VMs exhibit complementary temporal patterns, which can be leveraged to improve the oversubscription of underutilized resources.

Based on these insights, we propose \pname{}: a system that exploits temporal patterns for all-resource oversubscription in cloud platforms.
\pname{} uses long-term predictions and an efficient VM scheduling policy to exploit temporally complementary patterns.
We introduce a new general-purpose VM type, called \implvm{}, where we partition each resource allocation into a guaranteed and an oversubscribed portion.
\pname{} monitors the oversubscribed resources to detect contention and mitigate any potential performance degradation.
We focus on memory management, which is particularly challenging due to memory's sensitivity to contention and the overhead required to reassign it between \implvm{}s. 
Our experiments show that \pname{} enables platforms to host up to $\sim$26\% more VMs with minimal performance degradation.
\end{abstract}

%% file: intro.tex
%\vspace{-5pt}
\section{Introduction}
\label{sec:intro}
%\vspace{-4pt}

\myparagraph{Motivation}
Cloud platforms such as Microsoft Azure, Amazon Web Services (AWS), and Google Cloud Platform (GCP) offer compute resources (\eg, CPU, memory, and network) as virtual machines (VMs).
To meet the ever-increasing performance requirements of users, cloud providers are pressured to offer their services efficiently. 
However, achieving optimal efficiency remains challenging, and resource utilization in cloud platforms is often low~\cite{Mckinsey-2008, worth-HPCA2010, google-2012cc, datacenter-2022, quasar-2014, long-2014}.

There are three common causes of low resource utilization in cloud platforms.
First, platforms leave \emph{unallocated} resources for future VM allocations to ensure an optimal experience for customer workloads (\eg{}, rapid scale-out, high availability, and reliability) and for platform management (\eg{}, datacenter tax~\cite{seemakhupt2023cloud,kanev2015profiling}).
Prior work characterized unallocated resources and proposed solutions to minimize them, including Spot, Burstable, and Harvest VMs~\cite{harvestvm:osdi2020, burstablevm:aws, burstablevm:azure, spotvm:azure, preemptiblevm:google, mhvm:asplos2022, blockflex:osdi2022}.

Second, \emph{stranded} resources are unallocated but cannot be used to allocate new VMs because another resource on the server is fully allocated.
Prior work focused on memory stranding~\cite{pond:asplos2023} and proposed mitigating it using disaggregation~\cite{pond:asplos2023,faasmem:asplos2024}. 
While disaggregation is promising for memory, it is unavailable for other resources (\eg, CPU and network).

Third, \emph{underutilized} resources are allocated but not always used by the workload on the VM.
Prior work addressed this issue by using Harvest VMs, which can borrow underutilized resources from colocated VMs~\cite{harvestvm:osdi2020,mhvm:asplos2022,blockflex:osdi2022}. 
Unfortunately, users must modify workloads on Harvest VMs to account for evictions and dynamic resource allocations. 
Alternatively, using oversubscription can reduce underutilization by allocating fewer resources and multiplexing them between VMs on demand~\cite{borg-2020,heracles-isca2015,history:osdi2016, rc:sosp2017,poweroversub:atc2021,twine:osdi2020,caspian:amazon2023,ECS:amazon2019,mesos-nsdi2011,alibaba:iwqos2019,overdriver-vee2011,oversubpower-asplos2020, oversubstorage:vmware}.
However, it has not been applied holistically to cloud VMs.

To understand the potential of oversubscription for cloud VMs,
we study the resource utilization of over one million opaque VMs in Azure.
We observe: 
(1) large and long-running VMs consume the most resources;
(2) while CPU is usually the most underutilized resource~\cite{audible-asplos2024, rc:sosp2017, smartharvest:eurosys2021}, oversubscribing CPU can move the bottleneck for new VM allocations to other resources (\eg, memory and network), making holistic management of all resources essential;
(3) many VMs exhibit complementary temporal patterns (\eg, some have peak utilization at noon while others peak at night); and %, \hlgreen{which can be exploited to allocate fewer resources if the oversubscribed VMs are colocated}; and
(4) these patterns are predictable,
Cloud platforms can leverage these patterns to help colocate oversubscribed VMs~\cite{oversubamazon,oversubcpu:google}.

\myparagraph{Challenges}
Cloud users run highly heterogeneous workloads on opaque VMs.
Platforms must uphold strict service level objectives (SLOs) despite limited insights into diverse workload characteristics (\eg{}, tail latency sensitivity) and restricted telemetry visibility (\eg{}, CPU utilization).
The virtualization abstraction introduces further challenges. 
The resource management granularity is typically coarser compared to that of processes or containers, which have been the focus of many prior works~\cite{tmts:asplos2023,sdfm:asplos2019,tmo:asplos2022}.
Oversubscription should also be compatible with optimizations and platform management techniques.
Accordingly, we need to provide a transparent, safe, and resilient solution.

\myparagraph{Our work}
We propose \pname{}: a system to oversubscribe all VM resources.
\pname{} leverages temporal patterns in resource utilization to predict which VMs will have higher resource demands at specific times of the day.
Our scheduling policy identifies VMs with complementary resource utilization patterns for colocation.
Our policy takes a holistic approach to considering all resources.
\pname{} uses a new general-purpose VM type, called \emph{\implvm{}}, where each resource is partitioned into a guaranteed and an oversubscribed portion. 
The guaranteed portion is always allocated to the VM to maximize performance. 
In contrast, the oversubscribed portion is allocated on demand from an oversubscribed pool to maximize resource savings. 
Although we oversubscribe all resources, we focus on memory, one of the most sensitive and challenging, due to its non-fungibility.
\pname{} manages server resources and uses reactive and proactive mitigations to minimize the potential performance degradation caused by contention. 
\implvm{}s can be opt-in and discounted to compensate for the risk of performance degradation.

\myparagraph{Results}
We evaluate \pname{} using real workloads and production traces and quantify the trade-off between the resources saved and the risk of performance degradation.
We demonstrate that exploiting temporal patterns enables hosting up to $\sim$26\% more VMs with minimal platform overhead and VM performance degradation.

\begin{figure}[t]
    \centering
    \includegraphics[width=0.90\linewidth]{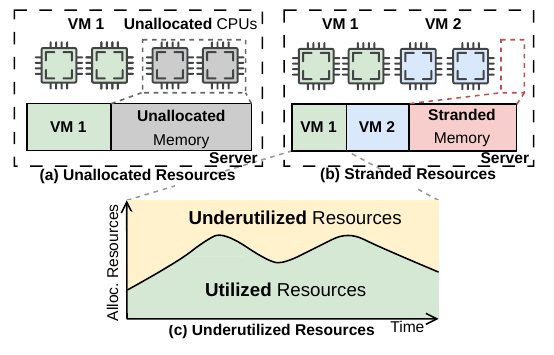}
    \vspace{-15pt}
    \caption{Examples of the causes of low resource utilization.}
    \label{fig:motiv_type_underutilization}
    \vspace{-5pt}
\end{figure}

\myparagraph{Summary}
We make the following main contributions:
\begin{itemize}[leftmargin=*]
\item Characterize the resource utilization of VMs in Azure, focusing on utilization over time and opportunities for oversubscription.
\item Propose \pname{} to oversubscribe all resources in cloud platforms and exploit temporal patterns at scale.
\item Introduce a new oversubscribed VM type, called \implvm{}, to ensure VM performance and maximize resource savings without requiring users to modify their workloads.
\item Quantify the trade-off of oversubscription between its savings and its potential impact on workload performance.
\end{itemize}

%% file: characterization.tex
% Characterization focusing on oversubscription

\section{Opportunity for oversubscription }
\label{sec:characterization}

Low resource utilization is ubiquitous in cloud platforms~\cite{worth-HPCA2010,google-2012cc,datacenter-2022,long-2014,overdriver-vee2011,alibaba:iwqos2019}.
It can be caused by resources that are:
(a) \emph{unallocated}, if they are unsold or reserved by the platform~\cite{rc:sosp2017,harvestvm:osdi2020,history:osdi2016,mhvm:asplos2022}; 
(b) \emph{stranded}, if they are unallocated but cannot be allocated due to the lack of other resources in the server~\cite{pond:asplos2023}; and 
(c) \emph{underutilized}, if they are allocated to a VM, but not used~\cite{smartharvest:eurosys2021,twine:osdi2020,ufo-nsdi2024,audible-asplos2024}.
\Cref{fig:motiv_type_underutilization} shows an example of each cause of low resource utilization. 
We study all three causes, focusing on their temporal patterns and the opportunities for oversubscription.

\myparagraph{Methodology}
We collected traces for two weeks in May 2024 of over one million opaque VMs from a subset of servers across ten popular clusters in seven Azure regions~\cite{rc:sosp2017}.
The traces cover thousands of servers from four hardware generations, including Intel and AMD processors.
For each VM, we record the allocation and deallocation times, resource allocation, server on which it runs, and maximum resource utilization for CPU, memory, network, and storage, respectively.
These utilization data are captured at 5-minute intervals (the default setting for long-term storage).
Using 5-minute intervals establishes a lower bound for underutilization, as we use the maximum utilization in each interval.

\begin{comment}
\begin{table}[t]
   \centering
    \caption{Sample clusters and their characteristics. \inigo{I may need to anonimize this a little.}}
    \label{tab:clusters}
    \footnotesize
    \begin{tabular}{ccccr}
        \toprule
        Cluster & Region & Generation & Hardware & \# Nodes \\
        \midrule
        C1	& 2.1 & 2 & Intel & 2915 \\
        C2	& 2.2 & 2 & Intel & 1930 \\
        C3	& 1.1 & 2 & Intel & 2854 \\
        C4	& 1.1 & 4 & Intel & 1994 \\
        C5	& 1.1 & 3 & Intel & 2215 \\
        C6	& 3.1 & 1 & Intel & 713 \\
        C7	& 1.2 & 3 & Intel & 1484 \\
        C8	& 3.2 & 2 & Intel & 2382 \\
        C9	& 1.4 & 2 & Intel & 2866 \\
        C10	& 2.3 & 3 & AMD   & 760 \\
        \bottomrule
    \end{tabular}
  
\end{table}
\end{comment}

\subsection{Characterizing allocated resources}
\label{sec:char-allocated}

We study the characteristics of VMs that reserve the most \emph{resource\customtimes{}hours} (\ie{}, allocated resources weighted by time), as oversubscribing these VMs can yield the greatest benefit.

\begin{figure}[t]
    \centering
    \includegraphics[width=\linewidth]{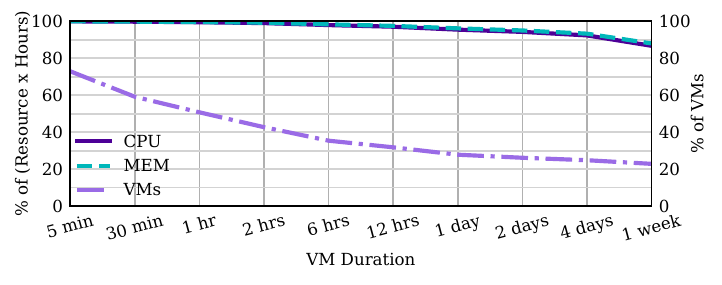}
    \vspace{-20pt}
    \caption{Percentage of resource\protect\customtimes{}hours consumed by VMs lasting longer than VM duration (left) and the percentage of VMs lasting more than VM duration (right).}
    \label{fig:motiv_mem_cpu_long}
    \vspace{-5pt}
\end{figure}

\myparagraphnodot{How long are resources allocated?}
\Cref{fig:motiv_mem_cpu_long} shows that VMs lasting more than one day consume $\sim$96\% of allocated cores\customtimes{}hours, despite accounting for only 28\% of the VMs.
This is also the case for memory (96\% of GB\customtimes{}hours), as most VMs have a similar ratio of memory to cores.
Network and storage show the same patterns.
These findings are consistent with those reported for CPU in prior work~\cite{rc:sosp2017}.

This observation shows an important distinction between the number of VMs and the resources they consume over time.
For example, sixty 32GB VMs lasting one minute and one 32GB VM lasting one hour both consume 32GB\customtimes{}hours.
Given this, we focus on VMs that last longer than one day.

\begin{figure}[t]
    \centering
    \includegraphics[width=\linewidth]{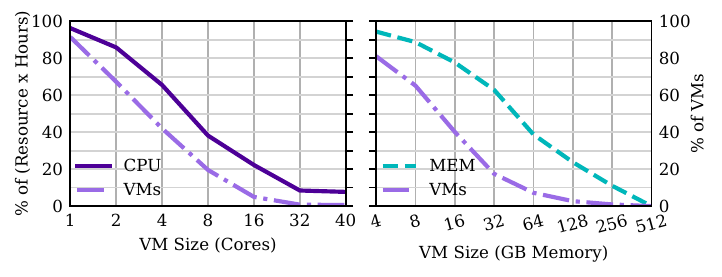}
    \vspace{-23pt}
    \caption{Resource\protect\customtimes{}hours and number of VMs consumed by VMs larger than a size (cores and memory).}
    \label{fig:motiv_vm_size_vs_resources}
    %\vspace{-5pt}
\end{figure}

\myparagraphnodot{What is the size of the VMs?}
\Cref{fig:motiv_vm_size_vs_resources} shows that larger VMs use more resources. 
The median VM in our study has 4 cores\footnote{We normalize hyperthreaded and non-hyperthreaded vCPUs to ``cores''.} and less than 16GB of memory, which is larger than indicated by prior work~\cite{rc:sosp2017}, where most allocated VMs had less than 2 cores and 4GB.
We observe the same distinction between the number of VMs and resources consumed as above.
For example, VMs with 32GB or more consume over 60\% of GB\customtimes{}hours, despite representing only $\sim$20\% of VMs.
Therefore, a solution targeting low resource utilization should account for both longer-running and larger VMs.

\subsection{Characterizing stranded resources}
\label{sec:char-stranded}

Cloud platforms can improve their packing of VMs into servers by aligning the resource ratios (\eg, GB/core) of VMs with the server hardware.
However, with the explosion of VM configurations~\cite{yadwadkar2017selecting,vmsizes:aws,vmsizes:google,vmsizes:alibaba} (\eg{}, 5 resource ratios, 9 sizes, 6 generations, and 4 specialized types in Azure~\cite{vmsizes:azure}), platforms may need to allocate VMs to servers with misaligned ratios.
For example, a server with hardware configured for general-purpose VMs (\eg, 4GB/core) may receive allocations for memory-optimized VMs (\eg, 16GB/core).
\Cref{fig:motiv_type_underutilization}b shows an example in which the cores are fully allocated, leaving memory stranded.
Prior work~\cite{pond:asplos2023} focused on stranded memory, while we focus on the stranding of all types of resources and the implications for oversubscription.

\begin{comment}
To use these stranded resources, prior work proposes oversubscription~\cite{oversubcpu:google, ECS:amazon2019, caspian:amazon2023, ufo-nsdi2024, audible-asplos2024, c2marl:www2023, shenango-nsdi2019, borg-2020} and ultimately disaggregation~\cite{pond:asplos2023,tpp:asplos2023,tmo:asplos2022,tmts:asplos2023, sdfm:asplos2019,vtmm:eurosys2023,slashing-eurosys2022}.
For example, placing a VM in a server that has not enough available memory but spare cores can use the memory in another server.
\end{comment}

\myparagraphnodot{How much are resources stranded?}
To study stranding, we place hypothetical VMs of the most typical VM configuration (\ie{}, 4GB/core)~\cite{d_series} on each server in our trace until one resource is exhausted and no further VMs can be placed. 
The remaining unallocated resources are considered stranded.
We repeat this calculation for each timestamp in the trace.

\begin{figure}[t]
    \centering
    \includegraphics[width=\linewidth]{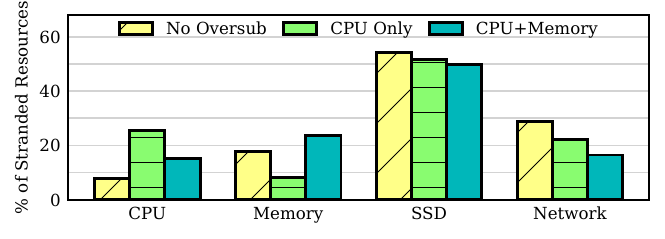}
    \vspace{-15pt}
    \caption{Average stranding for different resource types with varying levels of hypothetical oversubscription.
    }
    \vspace{-5pt}
    \label{fig:motiv_strand_mag}
\end{figure}

\textsc{No Oversub} in \Cref{fig:motiv_strand_mag} shows that CPU is the least stranded resource, with only 8\% stranding on average.
Memory, network, and SSD have 18\%, 29\%, and 54\% stranding, respectively.
Our analysis indicates that stranding may be more severe than previously reported~\cite{pond:asplos2023} and includes all types of resources.

\begin{figure}[t]
    \centering
    \includegraphics[width=\linewidth]{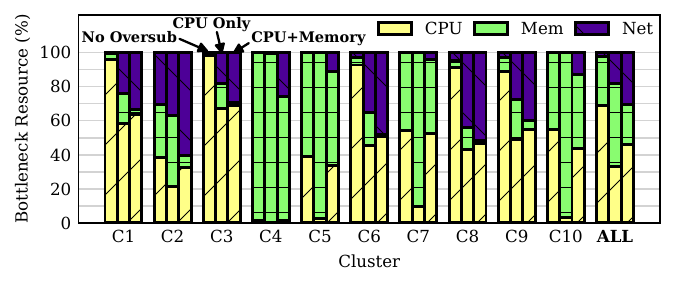}
    \vspace{-20pt}
    \caption{Percentage of time each resource is the bottleneck.
    }
    \label{fig:motiv_strand_bneck}
    \vspace{-5pt}
\end{figure}

\myparagraphnodot{Is stranding consistent across clusters?}
\Cref{fig:motiv_strand_bneck} shows the percentage of time each resource is the bottleneck for new VM allocations on a server (\ie{}, the cause of stranding) across all clusters.
\textsc{No Oversub} shows that the most common bottleneck is CPU, then memory, and finally, network.
This is inversely proportional to the amount of stranding shown in \Cref{fig:motiv_strand_mag}, as the bottleneck resource is fully allocated (\ie{} not stranded) on the server.
We omit SSD since it causes stranding less than 1\% of the time in all configurations.

We observe significant variation between clusters.
C1 is almost exclusively bottlenecked by CPU, C4 by memory, and C2 is divided between CPU, memory, and network.
This is because different clusters have different hardware configurations. 
For example, servers in C4 have less memory relative to cores/network than the other clusters.
Therefore, we need to account for the diverse configurations across servers.

\myparagraphnodot{What if we oversubscribe?}
Oversubscribing the bottleneck resource could unlock stranded resources for allocation.
\Cref{fig:motiv_strand_mag} shows the hypothetical impact of oversubscribing CPU (and memory) on stranding. 
We compute this by placing hypothetical VMs, as before, except that we also use underutilized CPU (and memory) resources to allocate these VMs.
For \textsc{CPU Only}, stranding increases for CPU to 25\% and decreases for memory, SSD, and network to 8\%, 52\%, and 22\%, respectively.
CPU stranding increases because some previously underutilized cores (\ie{}, allocated but unused) are now unallocated but bottlenecked by another resource.
We confirm this in \Cref{fig:motiv_strand_bneck}, where
the bottleneck shifts from CPU (69\% to 33\%) to memory (29\% to 49\%) and network (2\% to 18\%). 
We observe a similar trend for \textsc{CPU+Mem}: stranding for CPU, storage, and network decreases to 15\%, 50\%, and 16\%, respectively, while it increases to 24\% for memory.
Meanwhile, the bottleneck shifts from memory (49\% to 23\%).
In all cases, the utilization of each resource improves.
This motivates the need to oversubscribe resources holistically. %, as oversubscribed resources may be unutilized due to another bottleneck.}

\subsection{Characterizing underutilized resources}
\label{sec:char-underutilized}
Users may buy VMs with more resources than necessary because of variable load, performance sensitivity~\cite{smartharvest:eurosys2021,meng2010efficient}, or a mismatch with available options~\cite{yadwadkar2017selecting,alipourfard2017cherrypick}.
For example, \Cref{fig:motiv_type_underutilization}c shows a VM with an additional buffer to absorb workload bursts.
Alternatively, a user may need 6 cores and 12GB of memory, but the closest VM has 8 cores and 16GB.

\begin{figure}[t]
    \centering
    \includegraphics[width=0.49\linewidth]{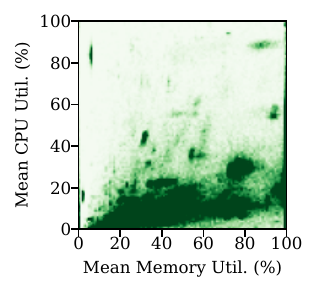}
    \includegraphics[width=0.49\linewidth]{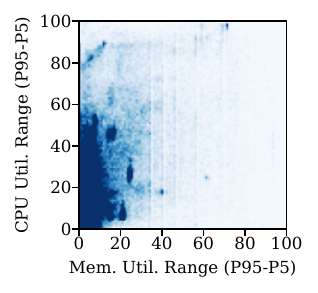}
    \vspace{-5pt}
    \caption{Correlation between CPU and memory for average utilization and utilization range across all VMs.
    }
    %\vspace{-15pt}
    \label{fig:motiv_cpu_vs_mem}
\end{figure}

Prior work characterized the underutilization of cloud resources, including CPU~\cite{rc:sosp2017,harvestvm:osdi2020,smartharvest:eurosys2021,history:osdi2016, twine:osdi2020,ufo-nsdi2024,audible-asplos2024}, memory~\cite{mhvm:asplos2022,borg:eurosys2015,borg-2020, compvm-2018}, SSD~\cite{history:osdi2016,blockflex:osdi2022,oversubstorage:vmware}, and power~\cite{poweroversub:atc2021,smartoclock-isca2024}.
We study the underutilization, correlations across different types of resources, temporal patterns, and implications for oversubscription.
We focus on VMs lasting over one day. %(\mbox{\Cref{sec:char-allocated}}).} 

\myparagraphnodot{What is the average utilization?}
\Cref{fig:motiv_cpu_vs_mem} shows the correlation between CPU and memory utilization.
The left half indicates that most VMs have an average CPU utilization below 50\%, which is consistent with prior works~\cite{rc:sosp2017,smartharvest:eurosys2021,ECS:amazon2019, heracles-isca2015, ufo-nsdi2024,alibaba:iwqos2019}, 
while there is greater diversity in the average memory utilization.
VMs with high CPU utilization also tend to have higher memory utilization.
The network and storage behavior resembles that of CPU.

\begin{comment}
\insightparagraph{6}
CPU is mostly underutilized while memory has a broader utilization spectrum.
\end{comment}

\myparagraphnodot{Does utilization vary?}
We define the utilization range as the difference between utilizations (\eg{}, P95-P5) over the lifetime of a VM.
The right half of \Cref{fig:motiv_cpu_vs_mem} shows that the range for CPU often reaches 60\%, while the memory is within 30\%, indicating that CPU utilization fluctuates more than that of memory.
In addition, 50\% of VMs have a memory range less than 10\%, and only 10\% of VMs have a range exceeding 50\%.
While VMs may have unique memory utilization, it typically fluctuates within narrow bounds.
The utilization patterns of network and storage resemble those of memory.

\myparagraphnodot{Is there a relation between average and range?}
We also measure the correlation between average utilization and range.
VMs with higher average CPU utilization tend to have a higher range.
Conversely, memory has a shorter range (less than 30\%) across all average utilizations.

\begin{figure}[t]
    \centering
    \includegraphics[width=\linewidth]{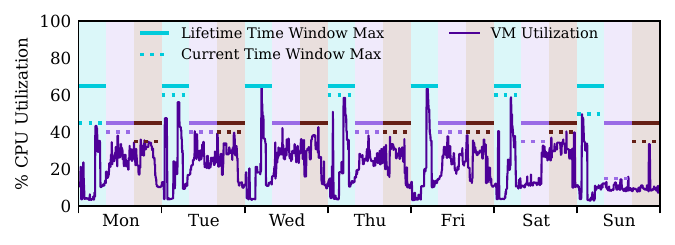}
    \vspace{-20pt}
    \caption{CPU utilization for a VM over a week split into three daily time windows: 0-8hr, 8-16hr, and 16-24hr.
    }
    \vspace{-5pt}
    \label{fig:motiv_vm_buckets_ex}
\end{figure}

\myparagraphnodot{Are there busier times in a day?}
Prior work classified the utilization patterns of bare-metal servers as periodic, constant, or unpredictable~\cite{history:osdi2016}.
Our work identifies peak times during the day (\eg, higher utilization every day at noon).
\Cref{fig:motiv_vm_buckets_ex} shows the CPU utilization of a VM with consistent daily peaks over a week.
We divide each day into three 8-hour windows: 0-8hr, 8-16hr, and 16-24hr.
For each time window, we show its peak utilization (current time window max) and its peak across the seven days (lifetime time window max),
rounded to 5\% buckets (\eg, 17.3$\rightarrow$20.0\%).
In the first window (0-8hr), the utilization is primarily under 10\% but has spikes up to 65\%.  %.
The remainder of the day (8-24hr) uses $\sim$40\% CPU. 
The current max is typically similar across days and close to the lifetime max. 

\begin{figure}[t]
    \centering
    \includegraphics[width=\linewidth]{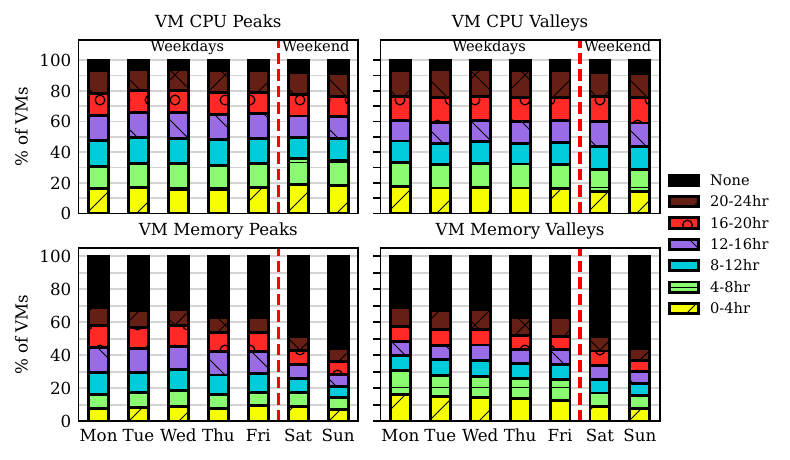}
    \vspace{-20pt}
    \caption{VMs with a peak/valley in each of the six 4-hour time windows for one cluster.
    }
    \label{fig:motiv_peaks_day}
    \vspace{-5pt}
\end{figure}

We first characterize the peak times for long-running VMs. 
A VM has a peak (and valley) in a given day if the difference between the maximum utilization in different time windows that day is at least 5\%. 
Any time window with a maximum utilization equal to the maximum (or minimum) across all time windows that day is counted as a peak (or valley).
Accordingly, a VM can have multiple peaks and valleys per day in different time windows. 
\Cref{fig:motiv_peaks_day} shows the percentage of VMs with peaks (and valleys) in each of six 4-hour time windows (\ie, 0-4hr, 4-8hr,\ldots), and those without peaks (\textsc{None}).
For each day, we normalize the total VMs with a peak (or valley) in each time window against the total VMs with a peak (or valley) that day.
Both CPU peaks and valleys are evenly distributed across the six time windows. Less than 10\% of VMs have no CPU peaks/valleys (\ie, their utilization is within a 5\% bucket). Nearly 70\% of VMs have memory peaks/valleys evenly distributed over time. 
This indicates that we could exploit these peaks and valleys by placing VMs that peak in specific time windows alongside VMs that have a valley at the same time.

\begin{figure}[t]
    \centering
    \includegraphics[width=\linewidth]{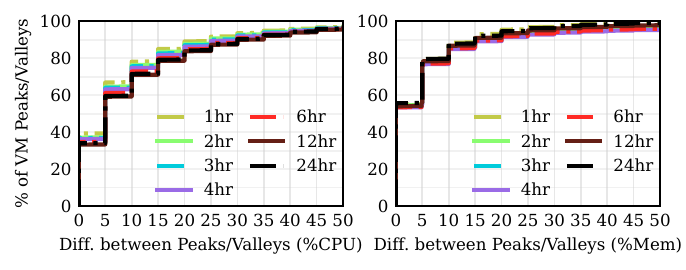}
    \vspace{-20pt}
    \caption{Difference in peak/valley utilization in consecutive days for different time window lengths.
    }
    \label{fig:motiv_all_bucket_next_day}
    \vspace{-5pt}
\end{figure}

\myparagraphnodot{Is the behavior consistent over time?}
\Cref{fig:motiv_all_bucket_next_day} shows the variation in peak and valley utilization in consecutive days. %for long-running VMs.
With 4$\times$6-hour windows, 80\% of VMs have a utilization difference of at most 20\% for CPU and at most 5\% for memory.
Overall, most VMs have consistent peaks and valleys, indicating that such patterns could be exploited over time.

\begin{figure}[t]
    \centering
    \includegraphics[width=\linewidth]{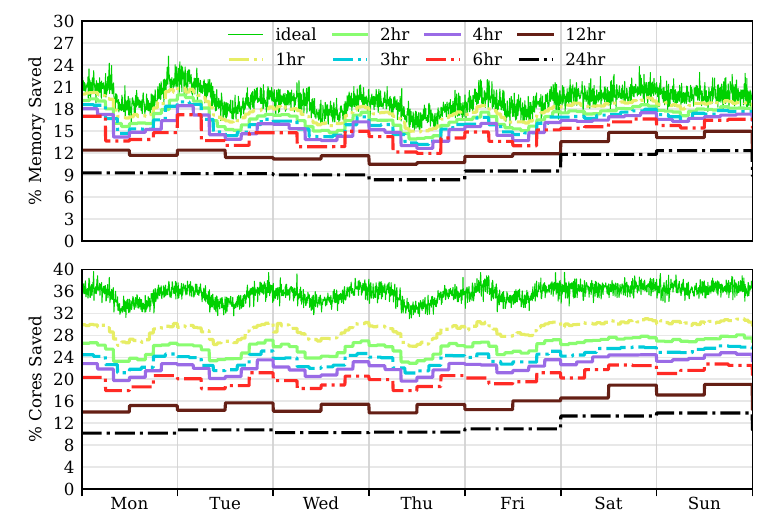}
    \vspace{-20pt}
    \caption{Potential savings for memory and CPU using time windows of multiple lengths for one cluster.}
    \label{fig:motiv_bucket_benefit}
    %\vspace{-5pt}
\end{figure}

\myparagraphnodot{Are patterns complementary?}
\Cref{fig:motiv_bucket_benefit} shows the percentage of allocated resources we can save in a representative cluster by packing VMs using their maximum utilization in each time window.
We compute the resources saved as the difference between oversubscription using these patterns (the maximum utilization in each time window) and overlooking them (the VM's lifetime max). 
For example, if a VM has a max utilization of 75\%, but its time windows have 30\%, 75\%, and 55\% utilization, we could save 45\%, 0\%, and 20\%, respectively.
We show the average savings across all VMs.

With a single 24-hour window, we save $\sim$8\% of memory and $\sim$8\% of CPU.
Using 4$\times$6hr windows, we save $\sim$15\% of memory and $\sim$20\% of CPU. %\hlgreen{while}
Multiplexing 5-min windows (ideal) saves $\sim$18\% memory and $\sim$34\% CPU.

\begin{figure}[t]
    \centering
    \includegraphics[width=\linewidth]{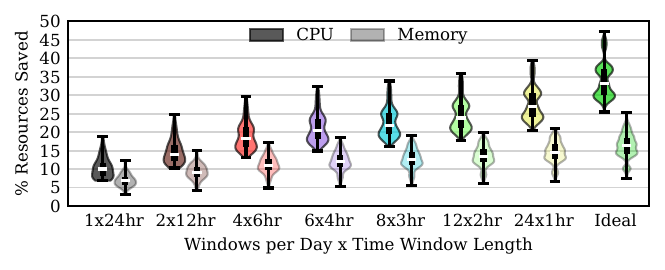}
    \vspace{-20pt}
    \caption{Summary of potential savings using different time windows across all clusters.  The distribution is shown as a colored violin, median by white lines, interquartile range (P75/P25) by black rectangles, and max/min by black lines. Darker/lighter violins are for CPU/memory, respectively.
    }
    \label{fig:motiv_bucket_benefit_summary}
    \vspace{-5pt}
\end{figure}

\Cref{fig:motiv_bucket_benefit_summary} summarizes the potential resource savings across all 10 clusters as a violin plot.
We observe that colocating VMs with complementary patterns can consistently result in significant savings across clusters.  
The savings increase with the number of windows but start plateauing with 6$\times$4-hour.
We can typically save more CPU than memory.

\begin{figure}[t]
    \centering
    \begin{subfigure}[t]{.49\linewidth}
    \centering
    \includegraphics[width=\linewidth]{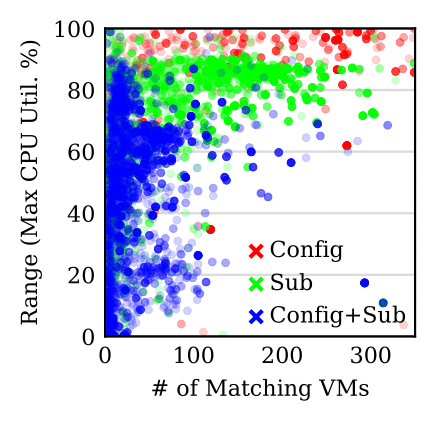}
    \vspace{-20pt}
    \caption{CPU.}
    \label{fig:va_pa_tradeoff_perf}
    \end{subfigure}%
    %\includegraphics[width=0.45\linewidth]{./Figures/motiv_pred_new_vms_cpu_scatter.png}%
    %\hspace{1.5pt}
    \begin{subfigure}[t]{.49\linewidth}
    \centering
    \includegraphics[width=\linewidth]{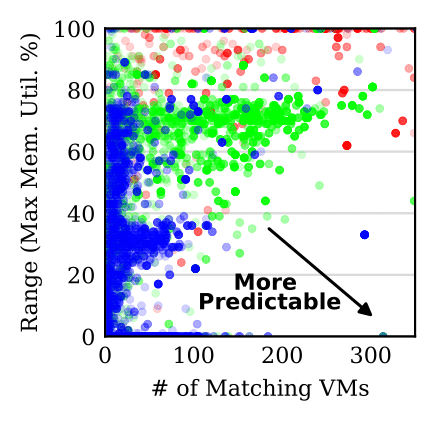}
    \vspace{-20pt}
    \caption{Memory.}
    \label{fig:va_pa_tradeoff_perf}
    \end{subfigure}%
    \vspace{-5pt}
    \caption{Per VM correlation between number of previous VMs of the same group and their utilization range. We use 3 groups: subscription, VM configuration, and a combination.
    }
    \vspace{-5pt}
    \label{fig:motiv_type_sub_vms}
\end{figure}

\myparagraphnodot{Are new VMs similar to old VMs?}
We analyze whether existing VMs can be grouped to use their aggregated resource utilization patterns to predict the patterns of future VMs. 
For each VM in the second week of our trace, we analyze the utilization of a group of similar VMs in the first week based on three groupings of similarity:
VMs from the same (1) customer subscription~\cite{rc:sosp2017},
(2) VM configuration, and
(3) subscription and VM configuration. 
Other features (\eg{}, VM name, guest OS version, or creation time) were less relevant. 
For each VM, \Cref{fig:motiv_type_sub_vms} shows the number of matching VMs from the same group (\eg{}, subscription) and the range of their maximum resource utilization.
For example, a VM with 10 prior VMs from the same subscription whose peak CPU utilizations were within a range of 10\% is plotted as $(10,10)$.
Ideally, we want many matching VMs (\eg{}, >50) with low ranges (\eg{}, <10\%). 

Grouping by VM configuration, the median VM has many previous VMs (over 2,000), but their utilization range is high (nearly 100 for memory).
Grouping by subscription, the median VM has fewer previous VMs (over 120), and their utilization range is smaller (under 70 for memory).
Grouping by both, the median VM has the fewest previous VMs (40) with the smallest range (only 31 for memory).
To determine predictability, we compare the maximum utilization of each VM with the average peak of its prior VMs.
With subscription and VM configuration, memory shows better predictability (over 70\% of VMs within 10\% of the average peak utilization) compared to CPU (70\% of VMs within 20\% of the average peak). 
Overall, most VMs have sufficient historical patterns that can aid in predicting future resource utilization.

%% file: design.tex
\section{Temporal pattern-based oversubscription}
\label{sec:design}

Based on \Cref{sec:characterization}, there is a significant opportunity to leverage complementary temporal utilization patterns for all resources due to their predictability.
Traditional resource oversubscription solutions~\cite{audible-asplos2024, cloudoversub:hotice2012, oversubstorage:vmware, twine:osdi2020, borg:eurosys2015} aimed to reduce underutilization in cloud platforms but did not exploit these patterns.
We propose \pname{}: a system to oversubscribe virtualized cloud platforms that leverages complementary temporal utilization patterns for all resources at scale.

\subsection{\pname{} overview}
\label{subsec:overview}

\begin{figure}[t]
    \centering
    \includegraphics[width=0.9\linewidth]{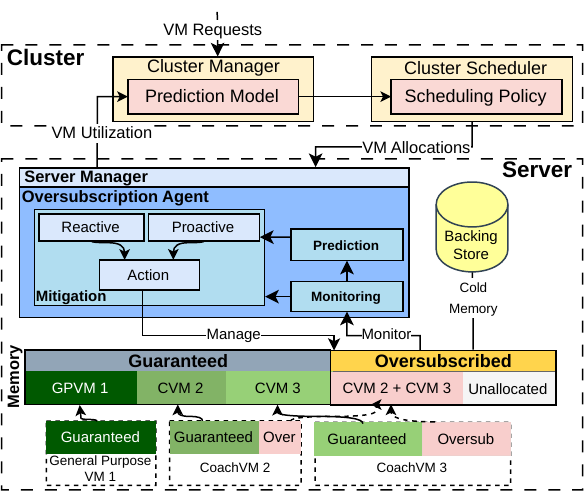}
    %\vspace{-13pt}
    \caption{Design overview of \pname{}.}
    %\vspace{-12pt}
    \label{fig:overview}
\end{figure}

We present the design overview of \pname{} in \Cref{fig:overview}, including the common workflow to create and operate oversubscribed VMs.
It comprises a logically centralized cluster management layer and a local component for each server.

\myparagraph{Cluster management}
When creating an oversubscribed VM~\cite{oversubamazon,oversubcpu:google}, the \emph{cluster manager} converts the request (\eg, 4 cores and 16GB of memory) into resource requirements and oversubscription rates.
It uses a \emph{prediction model} to decide the oversubscription rates for each resource in each time window
(\eg{}, oversubscribe memory by 30\% during the day and by 20\% at night).
The \emph{cluster manager} sends these rates to the \emph{cluster scheduler},
which uses them to assign the VM to a server and sends the request to the selected \emph{server manager}.

\myparagraph{Server management}
The local \emph{oversubscription agent} manages the resources on each server
and adjusts the guaranteed and oversubscribed resources whenever a VM is allocated or deallocated.
The agent consists of three components:
(1) \emph{monitoring} to collect utilization data;
(2) \emph{prediction}, to predict future utilization;
and (3) \emph{mitigation} to detect contention and take local (\eg, reassign resources) and global (\eg, migrate VMs) remediation actions.
The \emph{oversubscription agent} periodically sends the utilization data to the \emph{cluster manager} to improve prediction accuracy.

\myparagraph{Customer adoption}
Ideally, customers can seamlessly transfer as many workloads as possible to \pname{}. %to fully realize the potential for oversubscription.
To achieve this, we set the following additional design goals:
\begin{itemize}[leftmargin=*]
    \item \textbf{\goalmincust{}: Minimize customer burden.} \pname{} should be transparent to the workloads on the VM to allow customers to deploy unmodified workloads.  
    \item \textbf{\goalminworkl{}: Minimize workload interference.} \pname{} should minimize any negative impact on VM performance to maintain existing SLOs as much as possible. 
\end{itemize}

To achieve \goalmincust{}, \pname{} addresses the virtualization-specific challenges through \implvm{}s (\Cref{subsec:coachvm}) without requiring workload modifications from users. 
For \goalminworkl{}, \pname{} oversubscribes conservatively despite opportunities to save additional resources, thereby minimizing the chance of contention (\Cref{subsec:time_windows}). 
By default, \pname{} eschews techniques that rely on awareness of the workloads running in the VM.
However, such techniques can be integrated if desired.

\begin{figure}[t]
    \centering
    \includegraphics[width=\linewidth]{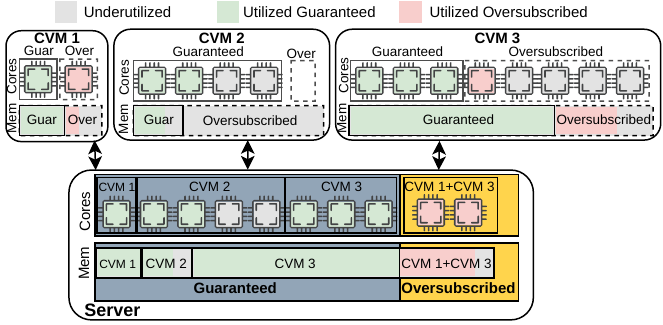}
    \vspace{-15pt}
    \caption{Three \implvm{}s in a server showing the guaranteed and oversubscribed CPU and memory.
    }
    \label{fig:hyvm}
\end{figure}

\subsection{\implvm{}s: Oversubscription for cloud VMs}
\label{subsec:coachvm}
Unlike containers, which are typically short-lived and share the host kernel, VMs face additional challenges for oversubscription. 
First, VMs have more limited telemetry visibility (\ie{}, they are opaque to the platform) and coarser resource granularity, complicating resource management.
Second, oversubscribed VMs must remain compatible with existing optimizations and platform management techniques (\eg{}, device assignment, live migration, and host updates). 

To support oversubscription, we introduce a new general-purpose VM type called \emph{\implvm{} (\cimplvm{})}.
It has a guaranteed portion of each resource for reliable performance and receives the remaining allocation on-demand from an oversubscribed pool for savings.
All resources are managed transparently to the VM, preserving its general-purpose nature.
This approach enables customers to run any guest OS and unmodified workloads without burdensome application changes, eliminating barriers to widespread adoption (\goalmincust{}).

\myparagraph{Guaranteed and oversubscribed}
\implvm{} resources are divided into \emph{guaranteed} (always allocated to the \cimplvm{} to ensure performance) and \emph{oversubscribed} (shared across \cimplvm{}s to save resources).
\Cref{fig:hyvm} shows how \pname{} may allocate resources to three \cimplvm{}s: % with different sizes:
\cimplvm{}1 with 2 cores and 8GB of memory, \cimplvm{}2 with 4 and 16GB, and \cimplvm{}3 with 8 and 32GB.
\pname{} guarantees 8 cores and 26GB of memory (\cimplvm{}1: 1 and 4GB, \cimplvm{}2: 4 and 4GB, and \cimplvm{}3: 3 and 18GB).
The remaining 6 cores and 30GB of memory are oversubscribed and backed by only 2 cores and 16GB.
In this way, \pname{} can fit VMs with 14 cores and 56GB of memory into a server with 10 cores and 36GB ($\sim$30\% oversubscription rate).

\begin{table}[t]
    \footnotesize
    \centering
    \caption{Common fungible and non-fungible resources and the mechanism used to share them across VMs.}
    \vspace{-5pt}
    \begin{tabular}{ccc}
        \toprule
         Resource             & Fungible & Mechanism\\
         \midrule
         CPU                  & \greencmark & CPU groups \\
         Memory space         & \redxmark   & PA/VA portions, VA-backing\\
         Memory bandwidth     & \greencmark & Shares, reservations, caps \\
         Network bandwidth    & \greencmark & Shares, reservations, caps \\
         Accelerated network  & \redxmark      & SR-IOV \\
         Storage bandwidth    & \greencmark & Shares, reservations, caps \\
         Local storage space  & \redxmark      & Disk partitions, DDA, SR-IOV \\
         Remote storage space & \greencmark & Cache size and network bandwidth \\
         GPU                  & \redxmark      & DDA, SR-IOV \\
         Power                & \greencmark & Frequency and power caps \\
         \bottomrule
    \end{tabular}
    \label{tab:fungible_resources}
    \vspace{-5pt}
\end{table}
\myparagraph{Fungible and non-fungible}
Certain resources are harder to share across \cimplvm{}s when oversubscribed. 
We consider resources that can be quickly reassigned between VMs as fungible~\cite{nu:ruan2023}.
For example, we can easily multiplex CPU and network bandwidth across multiple VMs.
In contrast, virtual memory pages are assigned to specific physical pages and need to be paged out before the physical page can be reassigned.
We consider these resources as non-fungible.
\Cref{tab:fungible_resources} summarizes the fungibility of common resources.

For fungible resources, we assign multiple VMs to the same resource and let the hypervisor quickly reassign them.
However, we must assign non-fungible resources carefully.
In all cases, we must monitor and mitigate potential contention.

\Cref{tab:fungible_resources} lists the mechanisms we use to manage common resources.
For example, we use CPU groups to assign a subset of cores statically and oversubscribe the rest.
For the rest of the paper, we focus on \emph{memory space}, as it is non-fungible and one of the most challenging to oversubscribe. 
The general techniques discussed can be applied to other resources.
We omit a thorough discussion of CPU as it has been extensively covered in prior work~\cite{smartharvest:eurosys2021, rc:sosp2017}.

\myparagraph{Memory oversubscription}
The memory space of VMs can be physically addressed (PA) or virtually addressed (VA).
The hypervisor allocates PA memory at VM creation time and statically maps it to the guest physical addresses (GPA).
It uses huge pages (1GB) to reduce TLB overheads.
Accessing PA pages provides high performance, but their static allocation limits flexibility, as they cannot be easily reassigned. 
Therefore, we use PA memory for the guaranteed portion.

The hypervisor manages VA memory using 1GB pages, which can be allocated/mapped to specific VMs at a smaller granularity on demand. 
It can be dynamically backed by a smaller amount of physical memory, using a backing store (\ie, disk) if it is insufficient.
Unused VA pages can be \emph{trimmed}/ unmapped, and 
\emph{paged in} from the backing store when accessed.  %\hlmag{using the host's page fault mechanism.}
Since VA pages may be unmapped, accessing VA memory may result in page faults, which is slower than accessing PA memory. 
Despite the risk of reduced performance, we use VA memory for the oversubscribed portion because it can be (de)allocated on demand.
\pname{} maps the oversubscribed (\ie{}, VA) portion to the VM's GPA as a NUMA node with no cores (zNUMA)~\cite{pond:asplos2023}, which funnels accesses to the guaranteed (\ie{}, PA) portion without guest changes.

Depending on the \emph{working set} of the VM (\ie{}, active pages in the GPA), we can adjust the PA/VA ratio.
For example, if the working set is typically below 16GB for a 32GB VM, \pname{} can allocate 16GB of PA memory and back the 16GB of VA memory with 8GB (25\% oversubscription).
This maximizes performance with the PA portion and resource savings with the VA portion.
However, selecting the optimal PA/VA ratio is crucial for balancing both goals.

\begin{figure}
    \centering
    \begin{subfigure}[t]{.5\linewidth}
        \centering
        \includegraphics[width=\linewidth]{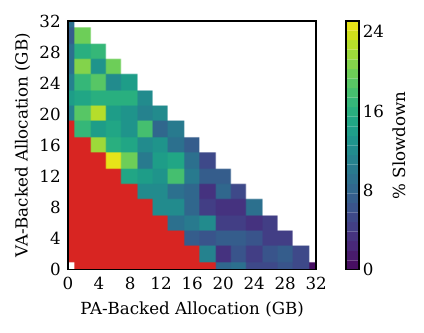}
        \vspace{-15pt}
        \caption{Performance slowdown.
        }
        \label{fig:va_pa_tradeoff_perf}
    \end{subfigure}%
    \begin{subfigure}[t]{.5\linewidth}
        \centering
        \includegraphics[width=\linewidth]{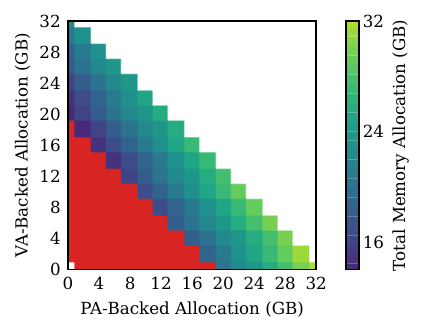}
        \vspace{-15pt}
        \caption{Memory allocation.}
        \label{fig:va_pa_heat_savings}
    \end{subfigure}
    \vspace{-5pt}
    \caption{Trade-off for PA/VA-backing with a 32GB \implvm{} that runs a workload with a working set of 18GB.
    In (b), we back 70\% of VA memory with physical memory.
    }
    %\vspace{-5pt}
    \label{fig:va_pa_tradeoffs}
\end{figure}

\myparagraph{Impact of PA/VA ratio on performance}
\Cref{fig:va_pa_tradeoff_perf} shows the performance impact on an unmodified memory-sensitive application with a working set of 18GB, running on a 32GB VM while varying the size of the PA and VA portions. 

The point with 32GB PA and 0GB VA (bottom right) is the baseline performance of a fully PA-backed VM (\ie, 0\% performance slowdown).
The white area represents invalid configurations (\ie{}, with more memory than the 32GB VM size or with no memory).
The red area represents configurations where the VM suffers unacceptable performance degradation due to continuous paging to disk.
The bottom right areas show minimal performance degradation.
This is because we can leverage the NUMA policies of unmodified guest OSes to transparently deprioritize the use of the VA portion with zNUMA. 
When we allocate less than 16GB of PA, we observe greater slowdowns.

\myparagraph{Impact of PA/VA ratio on memory savings}
\Cref{fig:va_pa_heat_savings} shows the total amount of allocated memory for the same 32GB VM with various sizes for the PA and VA portions.
We back only 70\% of the VA portion (based on the expected temporal-pattern multiplexing) and the entire PA portion with physical memory.
The fully PA-backed VM saves no memory, while a VM with 16GB PA and 16GB VA (backed by 12GB) saves 4GB. 
By combining the performance impact in \Cref{fig:va_pa_tradeoff_perf} and the savings in \Cref{fig:va_pa_heat_savings}, we can quantify the trade-offs when deciding the PA/VA ratio.

\myparagraph{Direct access to oversubscribed memory}
To expose resources such as GPUs, NVMe SSDs, and accelerated networking to VMs with high performance, cloud platforms use techniques like direct device assignment (DDA)~\cite{dda} (or device pass-through~\cite{kvm_passthrough}) and single-root input/output virtualization (SR-IOV)~\cite{msft_sriov,vmware_sriov}, which rely on direct memory access (DMA).
In VMs with oversubscribed memory, some parts of the GPA space may not be mapped to actual physical memory. When DMA requests attempt to access these unmapped memory regions, it can lead to I/O failures.
We discuss two solutions to address this issue. 

\myparagraphemph{Hardware support}
Devices with Address Translation Services/Page Request Interface (ATS/PRI)~\cite{ats_pri} can handle these intercepts.
ATS enables the device to use the second-level address translation table (SLAT) to translate GPAs and store translations locally.
PRI enables the device to handle SLAT failures (due to invalid pages) by requesting the host to fault the address as the CPU would. 
Therefore, these devices can access memory that could result in invalid translations. 

\myparagraphemph{Guest enlightenments}
Most devices do not yet support ATS/PRI.
In such cases, we need to ensure that any memory that may be accessed by a device is in the SLAT (\ie{}, has a valid mapping). %while the device} may be accessing that memory.
To address this, we introduce guest enlightenments (\ie{}, paravirtualization) that explicitly exchange memory ranges for I/O with the guest OS at boot time.
The host and guest then prevent moving or invalidating these ranges to avoid invalid translations.
\cimplvm{}s are also compatible with other existing guest enlightenments~\cite{enlightenments}.

\myparagraph{Compatibility with platform management}
Large-scale cloud platforms perform management operations to optimize VM packing and handle software/hardware maintenance.
To seamlessly deploy \pname{} at scale, \cimplvm{}s--particularly their VA memory--must be compatible with these operations.

\myparagraphemph{Live migration}
Cloud platforms use live migration to reduce fragmentation and perform maintenance~\cite{livemigrate:vee2018, vmphu:eurosys2021}.
Live migration already supports migrating PA-backed memory without modification. 
For the VA-backed portion, we must first page in any trimmed cold memory.
However, since copying cold memory happens during the pre-copy phase of migration~\cite{livemigrate:vee2018,vmphu:eurosys2021}, this does not extend VM downtime.

\myparagraphemph{Host updates}
Cloud platforms may need to update the host OS and reboot servers when there are critical security vulnerabilities~\cite{abu2019spectre,vmphu:eurosys2021}.
To perform these host OS updates, one could live migrate the VMs out of the server and reboot.
However, the overhead of doing this for every server in a data center is excessive.
To avoid this, platforms use VM preserving host updates, which temporarily pause the VMs and restore them after the host update~\cite{vmphu:eurosys2021}.
This host update preserves the VM memory through reboots.
This is simple for the PA-backed portion as it is directly mapped to the GPA.
In contrast, the VA portion relies on host OS memory management and involves complex data structures that are harder to persist across upgrades.
We incur this necessary complexity to persist these complex structures with negligible overhead.

\subsection{Utilization time windows}
\label{subsec:time_windows}

When placing a VM, platforms use a single allocation for the entire VM lifetime~\cite{protean:osdi2020}.
To exploit complementary patterns, \pname{} divides the VM utilization into time windows.
\Cref{fig:motiv_vm_buckets_ex} shows an example using three 8-hour windows per day.

\myparagraph{Predicting utilization}
To decide the oversubscription rate for \implvm{}s, we use a \emph{prediction model} that predicts the percentile (\eg, P95) utilization for each resource in each time window.
The model uses VM- and customer-specific features.
VM-specific features include the VM configuration, the weekday of allocation, and offering (PaaS vs. IaaS).
Intuitively, these inputs capture utilization behavior observed across customers. 
For example, utilization tends to be higher in VMs allocated on weekdays and IaaS VMs. 
Customer-specific inputs capture different utilization behavior between customers.  
We use the subscription type (\eg, internal production vs. test) and the history of resource utilization of previous VMs in that customer subscription.
The existing platform telemetry already collects all these inputs in the background, requiring no user input.
 
We use a random forest regressor~\cite{sklearn_rf} that predicts utilization in 5\% buckets (\eg, at most 65\%). %for each time window.
Random forest is well-suited for predicting VM utilization due to its effectiveness with categorical variables, as shown in prior work~\cite{harvestvm:osdi2020,pond:asplos2023}.
Among similar predictors (\eg{}, XGBoost~\cite{xgboost} and LightGBM~\cite{lightgbm}), we choose random forest because it tends to be less sensitive to overfitting. This can improve robustness and reduce the likelihood of underpredictions (\Cref{subsec:eval_perf}), minimizing the risk of worst-case contention (\goalminworkl{}).

To collect the training data, we aggregate the utilization data for each VM.
If there is insufficient data to predict a VM, we conservatively do not oversubscribe it.
We generate these predictions in the background to minimize overheads on the critical path of VM allocation.

\begin{comment}
\Cref{fig:pred} shows the full prediction workflow.
\begin{figure}
    \centering
    \includegraphics{}
    \caption{Prediction workflow. \todo{}}
    \label{fig:pred}
\end{figure}
\end{comment}

\myparagraph{VM scheduling policy}
Once \pname{} decides the oversubscription ratio for the \implvm{}, it picks which server to host the VM.
Traditional VM schedulers solve this bin-packing problem using heuristics that account for the availability of each resource~\cite{protean:osdi2020, schwarzkopf2013omega, ousterhout2013sparrow, boutin2014apollo}.
They use a vector with the VM requirements (\eg, \{8 cores, 32GB memory, 300GB SSD, 10Gbps\}) and check if it fits within the available server resources (\eg{}, \{16 cores, 24GB memory, 500GB SSD, 20Gbps\}).
For example, they cannot place this VM on this server due to insufficient memory.

\myparagraphemph{Scheduling time-windows}
Instead, \pname{} considers the predicted utilization of each resource for each time window. 
For fungible resources like CPU, we can simply use a vector of the predicted utilization for each time window.
For an 8 core \implvm{} with three time windows, we may predict 2 cores for 0-8hr, 6 for 8-16hr, and 4 for 16-24hr and try to place it in a server with 4 cores available for 0-8hr, 6 for 8-16hr, and 8 for 16-24hr.
Represented as vectors, we have \{2, 6, 4\} $\leq$ \{4, 6, 8\}, so the \implvm{} fits.

\begin{figure}[t]
    \centering
    \includegraphics[width=\linewidth]{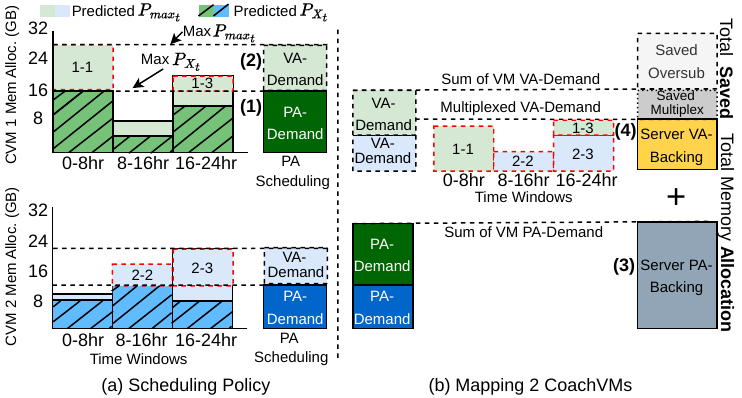}
    \vspace{-15pt}
    \caption{Memory allocation and mapping between time windows and PA/VA-backing.
    Two 32GB VM with 3 time windows map to 44GB (28GB PA and 16GB VA-backed).
    Numbers in parentheses refer to the formula that computes them.
    }
    %\vspace{-5pt}
    \label{fig:va_pa_mapping_2vms}
\end{figure}

\myparagraphemph{Non-fungible resources}
However, this approach does not consider if the resources can be reassigned easily (fungibility).
Specifically, the PA portion of the memory space cannot be easily adjusted at runtime (\ie{}, non-fungible).
To maximize performance, we would allocate PA memory for the VM's peak utilization across all time windows and not multiplex VMs with complementary patterns.
Conversely, using VA for the entire allocation to fully exploit temporal patterns may reduce performance.
Our approach is to balance these considerations.  
We maximize performance by allocating enough PA to satisfy the VM's working set a majority (\eg{}, 95\%) of the time.
We maximize multiplexing by allocating the remainder (\ie{}, up to the peak utilization) with VA. 

\Cref{fig:va_pa_mapping_2vms}a shows the VM scheduling for memory for two 32GB \implvm{}s with three time windows in a 48GB server.
For each window, we predict the maximum memory utilization (the total PA+VA-backed working set) and a percentile (\eg, P95) of the utilization for the guaranteed (PA-backed) portion.
The difference between the maximum and percentile represents the potential amount of VA-backed memory.  %, where accesses may experience degraded performance (\eg{}, using P95 could results in 5\% of accesses being degraded if accessed uniformly)}.

Initially, the scheduler ensures that the predicted maximum for each time window does not exceed the server's PA+VA-backed capacity:
$\{28,8,22\} + \{10,18,24\} \leq \{48,48,48\}$.
Next, since the PA portion is static across windows, the scheduler verifies the server has sufficient memory for the PA-backed portions: \texttt{16}+\texttt{12}<\texttt{48}.
Combining these checks into vectors, we have:
$\{28,8,22,\texttt{16}\} + \{10,18,24,\texttt{12}\}$ $\leq\{48,48,48,\texttt{48}\}$.
With this approach, the scheduler considers the number of windows plus one (for the max) for each resource, instead of just one, which incurs negligible overhead.

\myparagraph{Mapping time windows to \cimplvm{}s}
Once \pname{} assigns the \implvm{} to a server, we need to map the predicted utilizations to guaranteed and oversubscribed resources.
\Cref{fig:va_pa_mapping_2vms}b shows how we map the memory space for the \cimplvm{}s in \Cref{fig:va_pa_mapping_2vms}a. 
\pname{} allocates the maximum percentile prediction across all time windows for the guaranteed (\ie{}, PA) portion.
For the oversubscribed (\ie{}, VA) portion, one simple approach would be to allocate the sum of each VM's VA-demand.
However, this overlooks the fungibility of VA-backed memory and allocates excess memory.
Instead, we multiplex the VA-demand of each VM in each time window to further save memory 
by allocating the maximum multiplexed VA-demand. 
When the prediction component identifies potential changes in the resource requirements of a VM (\eg{}, across time windows), it notifies the mitigation component to reassign resources between \implvm{}s. 
Note that when assigning resources, we consider locality and relations between resources (\eg, NUMA). 
The server manager stores the VA-demand in each time window for each VM. It recomputes the multiplexed demand when it (de)allocates VMs and adjusts the oversubscribed portion accordingly. 

\myparagraphemph{Formulation}
$P_{\text{max}_t}$ and $P_{X_{t}}$ are the maximum and PX (\eg{}, P95) percentile for time window $t$, respectively:
%, and $VM_i$ a specific VM $i$.
%We have:
{\footnotesize
\begin{equation}
\forall i \in VM, \quad \text{PA\_demand VM}_{i} = \max_{t \in TW}(P_{X_{t}})
\end{equation}
\begin{equation}
\begin{split}
\forall t \in TW, \quad & \text{VA\_demand VM}_{i,t} \\
& = \max(0, ~P_{\max_{t}} - \text{PA\_demand VM}_{i})
\end{split}
\end{equation}
\begin{equation}
\text{Guaranteed memory} = \sum_{i \in \text{VM}} \text{PA\_demand VM}_{i}
\end{equation}
\begin{equation}
\text{Oversubscribed memory} = \max_{t \in TW} \left( \sum_{i \in \text{VM}} \text{VA\_demand VM}_{i,t} \right)
\end{equation}
}

\myparagraph{Choosing a prediction percentile}
We can navigate the trade-off between resource savings and potential performance impact by adjusting the prediction percentile.
For example, using P95 might risk 5\% of the accesses with lower performance but save 30\% of memory.

\begin{figure}
    \centering
    \begin{subfigure}[t]{.49\linewidth}
        \centering
        \includegraphics[width=\linewidth]{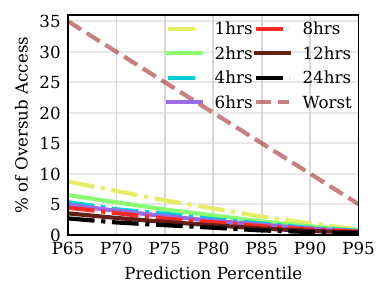}
        \vspace{-15pt}
        \caption{Oversubscribed accesses varying prediction percentile and time window.
        }
        \label{fig:va_pa_tradeoff_perf_2}
    \end{subfigure}%
    \begin{subfigure}[t]{.49\linewidth}
        \centering
        \includegraphics[width=\linewidth]{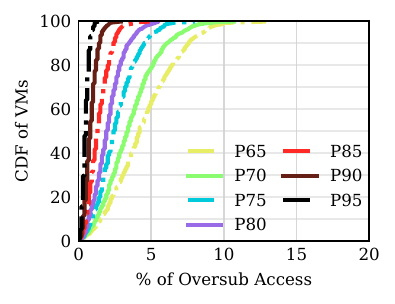}
        \vspace{-15pt}
        \caption{CDF of oversubscribed access percentage for 4hrs varying the prediction percentile.}
        \label{fig:va_access_tradeoff_2}
    \end{subfigure}
    \vspace{-5pt}
    \caption{Packing versus performance trade-off between PA and VA-backed memory for different time window lengths.
    }
    \vspace{-5pt}
    \label{fig:va_pa_tradeoffs_2}
\end{figure}

To choose the percentile, we estimate this trade-off using the VM traces from our study. 
\Cref{fig:va_pa_tradeoffs_2} shows the expected number of VA accesses based on the utilization percentile and the time window length, assuming each VM uniformly accesses its utilized memory.
\Cref{fig:va_pa_tradeoff_perf_2} shows that the VA accesses are much fewer than the prediction percentile for all time window lengths (\textsc{Worst}), due to rounding up the PA-allocation to 5\% buckets. 
For lower percentiles, the time window length is more important.
Finer-grained windows exploit more temporal patterns but risk additional accesses to oversubscribed memory.
\Cref{fig:va_access_tradeoff_2} shows the percentage of VMs with less than a percentage of VA accesses using a 4-hour window.
For example, 99\% of VMs have below 5\% VA accesses when predicting the P80 utilization. %utilization.

\Cref{fig:va_access_tradeoff_2} helps estimate additional accesses to oversubscribed memory that may result from under-allocating the guaranteed portion. % due to underpredicted resource utilization.
For example, allocating the P75 instead of P80 risks $\sim$1\% more accesses.
\pname{}'s scheduling policy is robust against such mispredictions.
Under-predictions only lead to under-allocation if they under-predict the maximum across all time windows, which may require multiple under-predictions.
Conversely, an over-prediction in a single time window can result in over-allocation.
While this may reduce savings, it is acceptable, as we prioritize protecting workload performance (\goalminworkl{}). 

\myparagraph{\pname{} configuration}
We configure \pname{} based on the performance vs. packing trade-off described above and time window length vs. packing trade-off in \Cref{fig:motiv_bucket_benefit_summary}.
To minimize the potential impact on VM workloads (\goalminworkl{}), we use the P95 utilization prediction and six 4-hour time windows.
This ensures oversubscribed resources are used at most 5\% of the time, though much less in practice.
We conservatively round allocations up to 5\% buckets and the resource management granularity (\eg{}, 1GB for memory).

\subsection{Monitoring and mitigating contention}
\label{subsec:monitor_mitig}

\myparagraph{Resource contention}
Oversubscribing resources introduces the risk of contention as the total utilized VM resources may exceed the available capacity. 
This can degrade the performance of VM workloads.
For CPU contention, VMs may need to wait for a specific core or run on other cores, which can hurt cache locality.
For memory contention, the hypervisor may page memory out, leading to increased disk I/O and latency.
\pname{} effectively reduces the potential impact of contention on VM workloads (\goalminworkl{}) by \emph{monitoring} and \emph{predicting} contention and \emph{mitigating} it when it occurs.
\pname{} uses generic metrics (\eg{}, CPU utilization) to monitor for contention without user input, but can be extended to include user metrics for further optimization (\eg{}, tail latency)~\cite{WI-2024}.

\myparagraph{Monitoring resource utilization}
The \emph{monitoring} component periodically (every 20 seconds in our implementation) tracks resource utilization and contention metrics (\eg, CPU wait time and page read/write operations).
To detect potential contention, it uses thresholds (\eg, >0.1\% CPU wait time at >20\% CPU utilization) computed using historical data at scale and correlated to performance incidents.
It notifies the \emph{mitigation} component to trigger reactive mitigations whenever it detects contention. 
We find that 20 seconds works well for memory, which spikes gradually.
For CPU, its fungibility can help reduce the impact of short spikes, allowing us to reduce the monitoring frequency. 
The \emph{monitoring} component sends this data to the \emph{prediction} and \emph{mitigation} components.

\myparagraph{Predicting contention}
The local \emph{prediction} component uses real-time and historical data to anticipate potential resource contention.
It uses a two-level prediction, an exponential weighted moving average (EWMA)~\cite{ingolfsson1993stability} that predicts the utilization for the next 20 seconds, and a long short-term memory network (LSTM)~\cite{yu2019review,pytorch_lstm} for the next 5 minutes.
EWMA is effective because resource behavior tends to be stable for short periods, while the LSTM better captures workload trends over time.
If the prediction exceeds the contention threshold, it notifies the \emph{mitigation} component to trigger proactive mitigations.
If \pname{} underpredicts the utilization, it relies on the \emph{monitoring} component to trigger reactive mitigations.

\myparagraph{Mitigating contention}
To prevent performance degradation, the \emph{mitigation} component triggers \emph{reactive} and \emph{proactive} mitigations when signaled by the \emph{monitoring} and \emph{prediction} components, respectively.

\myparagraphemph{Local mitigations}
For CPU, the local \emph{mitigation agent} first readjusts the CPU groups to meet actual demand.
Given that CPU is fungible, VMs under contention can borrow guaranteed idle cores from other VMs.
If necessary, \pname{} may increase the CPU frequency (if possible).
For memory, the agent first trims cold pages and, if necessary, requests the \emph{server manager} to add unallocated memory to the oversubscribed portion.
It uses similar techniques for other resources.

\myparagraphemph{Global mitigations}
When local mitigations are insufficient, the \emph{server manager} may ask the \emph{cluster manager} to evict lower-priority VMs (\eg, Spot VMs) or trigger live migration.
\pname{} decides which VM to migrate based on the potential to remedy contention (\eg, busier VMs cause more contention) and overhead (\eg, larger VMs require longer migration times).
As most resources are consumed by long-running VMs, \pname{} may migrate VMs whose resource requirements have changed. 
However, migration is the last option as it is the most expensive.

%% file: lessons.tex
%\section{Lessons from production}
%\subsection{Designing for production}
\subsection{Production considerations}
\label{sec:prod}

\myparagraph{Managing VMs vs. containers}
Unlike containers, VMs present additional challenges for cloud platforms due to their coarser resource granularity and opacity.
We developed the \implvm{} to address these transparently, requiring no workload or OS modifications.
Unlike short-lived, auto-scaled containers, VMs typically have more stable and longer temporal patterns.
\pname{} leverages these patterns for efficient VM scheduling and uses hypervisor-level metrics to transparently monitor and mitigate the impact of oversubscription.

\myparagraph{Maintainability and simplicity}
We consider the complexity of maintaining and operating \pname{}. % during deployment.
While more sophisticated learning-based approaches could offer marginal improvements, their decisions are often harder to interpret, making them impractical~\cite{mao2019learning,mao2016resource,duan2016benchmarking}.
Earlier versions of \pname{} used more complex algorithms but later shifted to higher-level concepts (\eg{}, time windows) that operators can more easily understand when troubleshooting issues.

\myparagraph{Modularity and extensibility}
\pname{} accounts for components at multiple levels of the stack (\eg{}, hardware, hypervisor, OS, and management agents), each with distinct rollout and development cycles.
While agents can be developed iteratively, hardware requires multi-year cycles.
Therefore, \pname{} is extensible and considers backward compatibility (\eg{}, hardware without ATS/PRI and legacy VM types).

\myparagraph{Staged rollout}
We initially oversubscribed fungible resources (\eg{}, CPU of certain first-party workloads~\cite{rc:sosp2017} and power via capping~\cite{poweroversub:atc2021}) at lower rates.
But, other resources like memory quickly become the bottleneck (\Cref{sec:char-stranded}).
As confidence builds (\eg{}, in mitigation effectiveness), we will oversubscribe them and increase the oversubscription rate.

\myparagraph{Customer awareness}
\pname{} simplifies adoption (\goalmincust{}) and minimizes the impact on customers' workloads (\goalminworkl{}).
However, users often prefer to know if they are oversubscribed.
Thus, \pname{} can be exposed to users as an opt-in feature at a discount and/or restricted to first-party VMs.
Our study of internal workloads showed that over 13\% were not user-facing, and nearly 25\% were delay-tolerant, characteristics that are favorable for oversubscription~\cite{workloadstudy}.

%% file: impl.tex
\subsection{\pname{} implementation}

We build \pname{} based on the system for static CPU oversubscription described in the Resource Central paper~\cite{rc:sosp2017}.

\myparagraph{\implvm{}}
We extend our existing VM management code to initialize the guaranteed and oversubscribed portions for each resource.
We use the mechanisms described in \Cref{tab:fungible_resources}, which are already available in Windows Server 2025~\cite{ws2025} and Hyper-V.
For memory, we initialize the virtual NUMA topology at VM boot time.
All these mechanisms are transparent and do not require any modifications to the guest.

\myparagraph{Utilization time windows}
We extend Resource Central~\cite{rc:sosp2017} to predict the utilization for each resource for each time window (\eg{}, time bins of 4 hours) using existing 5-minute resource utilization telemetry.
We add this as a new resource to our rule-based VM allocator~\cite{protean:osdi2020}.
We extend the local VM manager to store the resource requirements for each \cimplvm{} in each time window and perform multiplexing across \cimplvm{}s.

\myparagraph{Monitoring and mitigating contention}
We also extend the monitoring, prediction, and mitigation components 
in the local VM manager (\Cref{subsec:monitor_mitig}) to support new resources.
The monitoring component runs every 20 seconds and monitors existing OS performance counters~\cite{perfmon}. 
It tracks memory utilization by monitoring the VM page accesses.

The prediction component produces short-term predictions every 20 seconds and longer-term ones for the next five minutes using the data from the monitoring component.
The EWMA is updated in each 20-second window with the preceding resource utilization using $\alpha=0.5$. 
The LSTM uses the maximum and average utilization in the five previous 5-minute windows as input and is also updated online. 
As LSTMs require more input data, we train the model for 24 hours before using its predictions.

The mitigation component reuses existing OS capabilities to trim and flush cold memory and resize the oversubscribed memory partition~\cite{hyperv:memory}.
After paging in cold memory, we reuse the existing Hyper-V live migration~\cite{livemigrationhyperv}. 
Similarly, we make no changes to the process of killing low-priority VMs. 

%% file: eval.tex
\section{Evaluation}
\label{sec:eval}

Our evaluation shows that \pname{}:
(1) introduces minimal performance degradation to workloads running in \implvm{}s; % (\Cref{subsec:eval_perf}); 
(2) enables platforms to host up to 26\% more VMs;
(3) alleviates worst-case resource contention through mitigations; and % (\Cref{subsec:eval_mitig}); and 
(4) introduces minimal overhead to the platform.% (\Cref{subsec:eval_overhead}).

\subsection{Experimental setup}
\label{subsec:eval_setting}

\begin{table}[t]
    \centering
    \footnotesize
    \caption{Evaluated cloud workloads. }
    \vspace{-5pt}
    \begin{tabular}{lll}
        \toprule
            \textbf{Workload} & \textbf{Description} & \textbf{Key metric}\\
            \midrule
            Cache            & Memcached read/writes~\cite{memcached}. & Tail Latency \\
            Database         & Queries on a SQL database~\cite{pond:asplos2023}. & Tail Latency\\
            Big Data         & Sorting with TeraSort~\cite{terasort}. & Run Time\\
            Web              & 3-tier web application~\cite{specjbb}. & Throughput\\
            KV-Store         & Querying a KV-store~\cite{pond:asplos2023}. & Tail Latency\\
            Graph            & Computing pagerank~\cite{hibench}. & Run Time\\
            Microservices    & Social network~\cite{deathstarbench}. & Tail Latency \\
            LLM-FT           & BERT LLM fine-tuning~\cite{huggingfacetutorial}. & Run Time\\
            Video Conf       & Video conference application~\cite{pond:asplos2023}. & Throughput \\
        \bottomrule
    \end{tabular}
    \label{tab:workloads}
    %\vspace{-5pt}
\end{table}

We evaluate the impact of \pname{} on VM performance and estimate the overhead it introduces by running VMs on a real production server.
We also assess the effectiveness of \pname{}'s scheduling policy and examine performance violations at scale through simulations.

\myparagraph{VM workloads}
We use unmodified applications representing common cloud workloads, summarized in \Cref{tab:workloads}.
They run on VMs with unmodified Windows Server 2019~\cite{ws2019} and Linux Ubuntu 22.04~\cite{ubuntu2204}.
Each workload measures performance with a different \emph{key metric}.
\textsc{Cache}, \textsc{Database}, \textsc{KV-Store}, and \textsc{Microservices} are workloads with real-time requirements and their key metric is the P99 tail latency. The remaining workloads have no strict real-time requirements and their key metrics are either run time or throughput.

\myparagraph{Server}
We use a production server with two NUMA nodes and 160 hyper-threaded Intel CPU cores running at 2.3GHz and 512GB of DRAM.
The page file is placed on a Dell P5600 NVMe SSD~\cite{dellp5600}.
We reserve 2 cores and 4GB of memory to run \pname{}.
For each experiment, we ensure isolation by using CPU groups and placing the PA/VA memory into separate memory partitions.

\myparagraph{Simulator}
Before fully deploying \pname{}, which takes multiple years, we assess its benefits at scale using simulations. 
This enables us to evaluate multiple scheduling policies using identical VM traces on clusters with diverse hardware configurations without the need for more complex testing methods (\eg{}, A/B testing or input mirroring). 

Our simulator assigns VMs to servers by executing the real production VM scheduler code~\cite{protean:osdi2020} on the production VM traces (\Cref{sec:characterization}).
We extended the simulator to support long-term predictions and time-window-based scheduling, as described in \Cref{subsec:time_windows}.  
It is validated and closely mimics the constraints and preferences in the real scheduler. 
Based on the VM placements of the simulator, we simulate the resource utilization for each server using the 5-minute data and estimate the contention.

\subsection{\implvm{} performance} % and costs}
\label{subsec:eval_perf}
\pname{} minimizes the performance degradation from oversubscription. 
To demonstrate this, we execute the workloads in \Cref{tab:workloads} with four VM configurations:
\textsc{\gpvm{}}, which is fully guaranteed with PA; %  GVM
\textsc{\ovm{}}, which is fully oversubscribed with VA; % OVM
\textsc{\cimplvml{}}, which uses \pname{} to generate the PA/VA split; and
\textsc{\cimplvml{}-Floor}, which emulates an under-allocation (by 1GB). 
We run each experiment five times and create a new VM for each run.
We report the median for the \emph{key metric} of the workload (\Cref{tab:workloads}), with error bars for the max/min.

\begin{figure}[t]
    \centering
    \includegraphics[width=\linewidth]{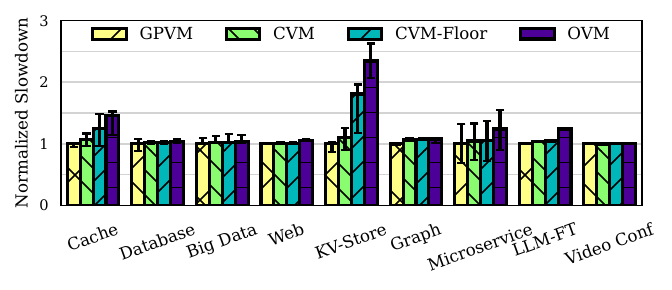}
    \vspace{-25pt}
    \caption{Performance of cloud workloads using various VM configurations.}
    \vspace{-10pt}
    \label{fig:eval_bench_perf}
\end{figure}

\myparagraph{Workload performance}
\Cref{fig:eval_bench_perf} shows the performance slowdown for each workload, normalized to the median of \textsc{\gpvm{}}. 
\textsc{Microservice}, \textsc{Cache}, and \textsc{KV-Store} are the most sensitive to oversubscription, with their performance degrading by up to 2.35$\times$ in the worst case (0.41 vs. 0.96ms) with \textsc{\ovm{}}. 
These workloads access memory in the critical path that may need to be allocated on demand, which degrades their tail latency.

The conservative allocations strategy from \pname{} and memory funneling with zNUMA prevent worst-case degradation (at most 10\%, from 0.41 to 0.45ms), as shown by \textsc{\cimplvml{}}. 
The other real-time workloads (\textsc{Cache}, \textsc{Database}, and \textsc{Microservice}) have 7\% (6.32 vs. 6.77ms), 2\% (40 vs. 41ms), and 4\% (2.71 vs. 2.83ms) tail latency slowdown, respectively.
This demonstrates that \pname{} can support even sensitive workloads with real-time requirements. 

Among workloads using other key metrics, \textsc{Llm-ft} is the most sensitive (1.24$\times$ worse: 3.7 vs. 4.5 mins) because it has the largest working set and frequently allocates/deallocates memory for each training iteration.
The limited memory reuse and frequent turnover stress the lower TLB reach and on-demand allocation, reducing performance. 
The remaining workloads experience at most 6\% slowdown with \textsc{\cimplvml{}}. 

\myparagraph{Performance with under-allocations}
Under-allocating the guaranteed portion can result in 1.8$\times$ performance degradation (0.41ms vs. 0.74ms), as shown by \textsc{\cimplvml{}-Floor}.
\textsc{KV-Store} and \textsc{Cache} are more sensitive to under-allocation than other workloads. 
Since their working sets are smaller, the oversubscribed portion receives a larger fraction of the total memory accesses. 
The remaining workloads experience at most an 8\% slowdown with \textsc{\cimplvm{}-Floor}. 
However, \pname{}'s scheduling policy is robust against under-allocations (\Cref{subsec:time_windows}), and rarely under-allocates VMs (\Cref{subsec:eval_cluster_util}).

\subsection{Impact of time window scheduling}
\label{subsec:eval_cluster_util}

\pname{} effectively predicts oversubscription rates to maximize available capacity and minimize resource contention. 

\begin{figure}
    \centering
    \begin{subfigure}[t]{.49\linewidth}
        \centering
        \includegraphics[width=\linewidth]{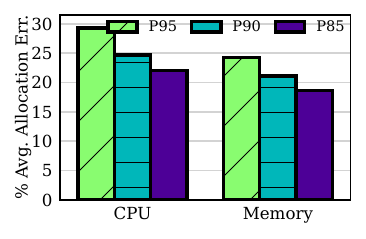}
        \vspace{-20pt}
        \caption{Over-allocation.}
        \label{fig:eval_sched_avg_err}
    \end{subfigure}
    \begin{subfigure}[t]{.49\linewidth}
        \centering
        \includegraphics[width=\linewidth]{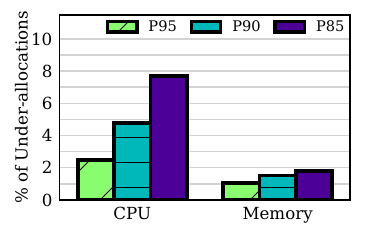}
        \vspace{-20pt}
        \caption{Under-allocations.}
        \label{fig:eval_sched_underpred_err}
    \end{subfigure}
    \vspace{-5pt}
    \caption{Effectiveness of \pname{}'s long-term predictions for different prediction percentiles. }
    \vspace{-15pt}
    \label{fig:eval_sched_err}
\end{figure}
\myparagraph{Prediction accuracy}
We evaluate the effectiveness of \pname{}'s predictions for the cluster-level, time-window-based VM scheduling policy in avoiding potential performance degradation from under-allocations while maximizing resource savings. % using different prediction percentiles.
\Cref{fig:eval_sched_avg_err} shows the over-allocation error using P95, P90, and P85 prediction percentiles.
The average error is 23-30\% for CPU and 19-24\% for memory.
The positive error is the additional resources that could have been saved compared with the ideal VM allocation. 
As we decrease the prediction percentile, the error decreases.

\Cref{fig:eval_sched_underpred_err} shows the under-allocations when allocating fewer resources than the ideal allocation.
Memory has few under-allocations (1-2\%), while CPU has a slightly higher percentage (3-8\%).
Predicting CPU is harder due to its higher fluctuations, but it is also more fungible, which reduces the impact of under-allocations.
Our scheduling policy helps mask the impact of individual under-predictions, as only an under-prediction reducing the maximum across all time windows results in an under-allocation.
Multiplexing \implvm{}s reduces this further.
This demonstrates that \pname{} effectively prioritizes minimized performance degradation (\ie{}, fewer under-allocations) for maximized resource savings (\ie{}, fewer over-allocations).

\begin{figure}
    \centering
    \begin{subfigure}[t]{.49\linewidth}
        \centering
        \includegraphics[width=\linewidth]{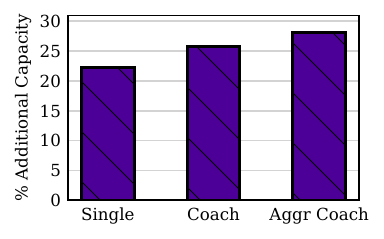}
        \vspace{-15pt}
        \caption{Additional capacity normalized against \textsc{None}.
        }
        \label{fig:eval_addtl_cap}
    \end{subfigure}
    \begin{subfigure}[t]{.49\linewidth}
        \centering
        \includegraphics[width=\linewidth]{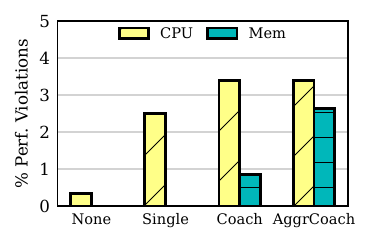}
        \vspace{-15pt}
        \caption{Performance violations.
        }
        \label{fig:eval_perf_vio}
    \end{subfigure}
    \vspace{-5pt}
    \caption{Impact of various oversubscription policies.
    }
    \vspace{-5pt}
    \label{fig:eval_sim_results}
\end{figure}

\myparagraph{Packing}
We evaluate the cluster-level savings brought by \pname{}'s time-window-based VM scheduling policy.
\Cref{fig:eval_addtl_cap} shows the additional sellable capacity (\ie{}, additional VMs that can be hosted) generated by different oversubscription policies. 
We simulate four policies.
(1) No oversubscription (\textsc{None}), which allocates all the VM's requested resources.
(2) Single oversubscription rate per VM (\textsc{Single}), which predicts a static oversubscription rate for each resource.
This represents a baseline similar to the state-of-the-art~\cite{borg-2020,twine:osdi2020,caspian:amazon2023,rc:sosp2017}. 
(3) Time-window-based oversubscription per VM (\textsc{\pname{}}). 
(4) An aggressive \pname{} (\textsc{Aggr \pname{}}), which uses a P50 prediction percentile.
\textsc{Single} increases capacity by 22\% compared to \textsc{None}.
\textsc{\pname{}} provides an additional 16\% capacity over \textsc{Single} and, 
\textsc{Aggr \pname{}} adds 9\% more capacity over \textsc{\pname{}}. \textsc{\pname{}} also reduces the number of required servers by 44\% through improved consolidation of VMs onto servers. 

\myparagraph{Performance}
We quantify how effectively \pname{}'s scheduling policy minimizes the resource contention from oversubscription.
CPU contention occurs when demand exceeds 50\% of the server capacity, while memory contention occurs when memory accesses result in page faults.
\Cref{fig:eval_perf_vio} summarizes the contention for each policy. 
\textsc{Single} adds 2\% CPU contention and no memory contention.
\textsc{\pname{}} increases CPU contention by 1\% while resulting in less than 1\% memory violations.
\textsc{Aggr \pname{}} introduces an additional 2\% memory violations.
Overall, \pname{}'s predictions and scheduling policy successfully leverage complementary temporal patterns to improve utilization with minimal impact.

\subsection{Mitigating contention}
\label{subsec:eval_mitig}

As oversubscription may still introduce some contention, we evaluate how effectively \pname{} identifies and mitigates it to minimize performance degradation.
We focus on memory in this subsection due to its sensitivity to contention. 

\myparagraph{Predicting contention}
The \textsc{EWMA} has low error due to the overall stability of memory utilization, with an average error under 4\% for over 85\% of VMs. 
The \textsc{LSTM} better captures historical utilization patterns to predict utilization and achieves only 2\% average error for 95\% of VMs.
The \textsc{LSTM} more accurately predicts VMs with dynamic but predictable patterns where the \textsc{EWMA} can show significant error. 
The combination of these two predictors allows \pname{} to effectively predict contention and trigger proactive mitigations.

\myparagraph{Mitigation policies}
To demonstrate the effectiveness of our mitigation policies in minimizing performance degradation,
we evaluate six policies against a baseline with no mitigation (\textsc{None}).
\textsc{Trim} only implements trimming.
If no cold memory is available for trimming, \textsc{Extend} may expand the oversubscribed memory pool with unallocated memory, and \textsc{Migrate} migrates a VM to free resources.
The \textsc{Reactive} variations trigger mitigation only after the monitoring component (\Cref{subsec:monitor_mitig}) detects contention, while the \textsc{Proactive} variations use the prediction component to proactively trigger mitigation.

\myparagraphemph{Effectiveness of mitigation policies on memory}
To demonstrate the effectiveness of \pname{}'s mitigation policies, we evaluate their impact on memory contention and workload performance.  
We use the two most memory-sensitive workloads, \textsc{Cache} and \textsc{KV-Store}, to show the worst-case scenario.  
We colocate them with a \cimplvm{} running \textsc{Video Conf}, which uses more memory than predicted, causing contention twice. 
Each workload runs on an 8GB \cimplvm{}. 
The working set for both \textsc{Cache} and \textsc{KV-Store} is $\sim$4GB, and they run on \cimplvm{}s with 3GB-PA (and 5GB-VA).
The working set of \textsc{Video Conf} is 5GB, but it runs on a \cimplvm{} with 1GB-PA (and 7GB-VA). 
We allocate an initial 6GB to the oversubscribed pool to back the 17GB of VA in the \cimplvm{}s.

\begin{figure}[t]
    \centering
    \begin{subfigure}[t]{0.99\linewidth}
        \centering
        \includegraphics[width=0.99\linewidth]{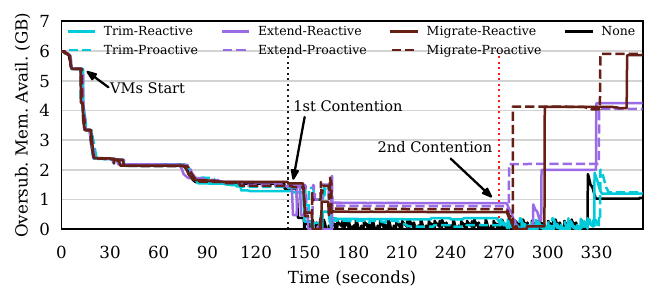}
        \vspace{-6pt}
        \caption{
        Available VA-backed memory during contention.}
        \label{fig:eval_mitig_memory}
    \end{subfigure}
    \begin{subfigure}[t]{0.99\linewidth}
        \centering
        \includegraphics[width=0.99\linewidth]{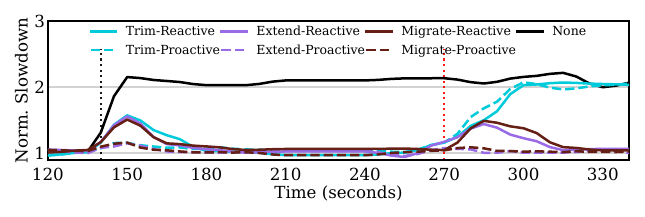}
        \vspace{-6pt}
        \caption{Zoom in on \textsc{Cache} performance during contention.}
        \label{fig:eval_mitig_cache}
    \end{subfigure}
    %\hfill
    \begin{subfigure}[t]{0.99\linewidth}
    \centering
        \includegraphics[width=0.99\linewidth]{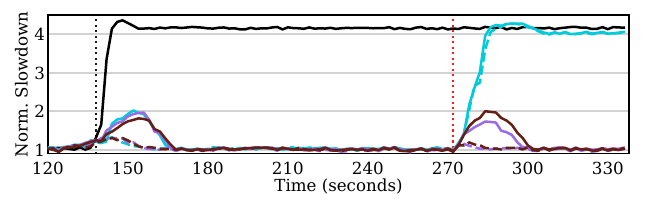}
        \vspace{-6pt}
        \caption{Zoom in on \textsc{KV-Store} performance during contention.}
        \label{fig:eval_mitig_kv}
    \end{subfigure}
    \vspace{-7pt}
    \caption{Comparison of \pname{}'s mitigation policies during two memory contentions.
    }
    \vspace{-15pt}
    \label{fig:eval_mitig_perf}
\end{figure}

\Cref{fig:eval_mitig_memory} shows the available memory in the oversubscribed portion (\ie, VA-backed).
The first contention starts at 135 seconds, when \textsc{Video Conf} consumes more memory than initially predicted.
For the first contention, VMs have enough cold memory that can be trimmed.
The second contention starts at 255 seconds, when \textsc{Video Conf} increases its working set, exceeding the amount of cold memory available for trimming.
The first contention demonstrates the effectiveness of trimming, while the second demonstrates the effectiveness of the other mitigation measures. 

\textsc{None} frequently pages out memory that is paged in later and fails to recover from contention.
\textsc{Trim} identifies large portions of cold memory in advance to reduce the number of trimming operations, and resolves the first contention quickly.
As the other policies also trim, they exhibit similar performance during the first contention but differ during the second.
\textsc{Trim} cannot recover from the second contention because there is insufficient cold memory. 
\textsc{Extend} quickly mitigates the second contention by expanding the oversubscribed pool with unallocated memory from the server. 
\textsc{Migrate} takes longer than \textsc{Extend} to resolve it as the memory cannot be reclaimed until \textsc{Video Conf} is migrated.
Across all policies, \textsc{Proactive} triggers mitigations earlier and resolves contention faster than \textsc{Reactive}. %, \hl{resulting in at most 30\% short-degradation for our most sensitive real-time workloads}.

\myparagraphemph{Impact of mitigation policies on performance}
\Cref{fig:eval_mitig_cache,fig:eval_mitig_kv} show the performance of the VMs during contention.
Contention degrades performance by up to 4.3$\times$, while our proactive policies reduce this overhead to only 1.3$\times$ by reducing the duration and intensity of the contention.
This demonstrates that \pname{}'s mitigation policies effectively minimize the performance degradation from contention. 
The other workloads experience less degradation.

\subsection{\pname{} platform overheads}
\label{subsec:eval_overhead}

\pname{} minimizes overhead by reusing and extending existing systems in Azure when possible. 
We demonstrate this by profiling the overhead \pname{} introduces to cloud platforms. 

\myparagraph{Predicting utilization time windows} 
To train the model %each data point consumes 16 bytes per time window and resource. 
for 6 time windows with about one million VMs,
\pname{} requires under 100MB of data and 121 seconds for daily offline training.
The model consumes 186MB of memory.

\myparagraph{Scheduling overheads}
From the simulations, the additional six dimensions for bin-packing each resource introduce less than 1ms to the scheduling time of a VM.

\myparagraph{\implvm{}s}
In \Cref{subsec:eval_perf}, the worst case page fault count for \textsc{\cimplvml{}} is less than 15\% of the ones for the \textsc{\ovm{}}.
In addition, the oversubscribed portion requires tracking accesses for trimming.
This requires 8MB of memory for a typical 32GB VM. 
Tracking every 20 seconds requires 2 additional hyper-threaded cores ($\sim$1.25\% overhead).
We offset this overhead by saving 16\% through CPU oversubscription (15\% net savings). 

\myparagraph{Predicting local contention}
Each local predictor requires 25KB of memory and 0.86ms for each training/inference cycle (every 5 minutes).
Even at peak load with >100 VMs on a single server, these overheads are easily absorbed by the existing cores reserved for server management.

\myparagraph{Mitigation measures}
\pname{} supports trimming and extending the resource pool to help alleviate memory contention. 
We can achieve a trim bandwidth of 1.1GB/s. %\hlgreen{as cold memory must be flushed to the backing store. }
Extending the oversubscribed pool achieves higher bandwidth at 15.7GB/s as it does not require writing cold memory to the backing store.

%% file: related.tex
\section{Related work}
\label{sec:related}

\myparagraph{Resource oversubscription}
Numerous studies have explored resource oversubscription~\cite{accelnet:nsdi2018, poweroversub:atc2021, ufo-nsdi2024, audible-asplos2024, c2marl:www2023, cloudoversub:hotice2012, overdriver-vee2011, livemigrate:vee2018,ECS:amazon2019}
and cloud providers already offer static oversubscription~\cite{borg:eurosys2015, borg-2020, autopilot-EuroSys20, heracles-isca2015, rc:sosp2017, history:osdi2016, autopilot-isard2007, history:osdi2016, twine:osdi2020, oversubamazon, caspian:amazon2023}.
Several studies have focused on optimizing individual resource utilization in data centers, including CPU~\cite{ufo-nsdi2024, audible-asplos2024, c2marl:www2023, shenango-nsdi2019}, memory~\cite{borg-2020, twine:osdi2020, heracles-isca2015, overdriver-vee2011, caspian:amazon2023}, power~\cite{poweroversub:atc2021, powercapping-2013need, powerstatistical-eurosys2009, smoothoperator-asplos2018, managingpower-sigarch2012, thunderbolt-osdi2020, ensemble-sigarch2006, oversubpower-asplos2020, dynamo-isca2016}, and storage~\cite{blockflex:osdi2022, oversubstorage:vmware}.
\pname{} is the first work to target all resources in a virtualized environment while exploiting complementary temporal patterns to improve utilization.

\myparagraph{Oversubscribed containers}
Borg~\cite{borg:eurosys2015, borg-2020} manages cluster scheduling and allocation, optimizing resource use through job oversubscription.
Google's Autopilot~\cite{autopilot-EuroSys20} configures task concurrency and CPU/memory limits using machine learning and historical data to reduce slack and task failures. 
Twine~\cite{twine:osdi2020,twineoversub:socc2024} orchestrates containers across servers using dynamic machine partitioning, preventing capacity stranding and allowing CPU or memory oversubscription upon user request. 
Twine SRM adjusts job tasks based on historical data. 
\pname{} introduces the \implvm{} to overcome the additional challenges for all-resource oversubscription in virtualized environments (\eg, opaqueness, live migration, direct device assignment, and host updates).

\myparagraph{Oversubscribed VMs}
EC2~\cite{caspian:amazon2023} offers VMs with flexible resource allocation and oversubscription capabilities, leveraging a shared Linux kernel to optimize CPU and memory utilization. %while requiring effective management to minimize performance issues.
\pname{} employs the new \implvm{}, which has a guaranteed and oversubscribed portion of each resource. This enables platforms to ensure performance while exploiting complementary temporal patterns to save additional resources without requiring users to modify their workloads. 

\myparagraph{Workload-aware scheduling}
Researchers leveraged learning techniques to optimize resource efficiency while ensuring SLOs~\cite{compvm-2018, consolidating-infocom2014, Wrasse-socc2012, cloudscale-cc2011, poweroversub:atc2021, bianchini2020toward, rlscheduler-sc2020,zhang2021flex, WI-2024} .
Some studies utilized historical information about the services for intelligent task scheduling and data placement~\cite{history:osdi2016, heracles-isca2015, twine:osdi2020, ubis:ipdps2018, mesos-nsdi2011}, while others concentrated on optimizing server and VM utilization, improving scalability and fault tolerance~\cite{smartoclock-isca2024, rc:sosp2017, li2017holistic, kim2009task}.
\pname{} uses generic metrics to predict and leverage patterns in resource utilization. 
In addition, it uses these metrics to predict and proactively mitigate contention, ensuring efficient utilization of resources and maximized workload performance without requiring workload awareness.

%% file: conclusion.tex
\section{Conclusion}
\label{sec:conclusion}

We introduced \pname{}, a system to improve resource utilization in cloud platforms by leveraging temporal patterns in VM workloads.
Our comprehensive characterization of resource utilization revealed that VMs often exhibit complementary patterns, which \pname{} exploits to increase oversubscription without compromising performance.

\pname{}’s time-window-based predictive scheduling policy enables cloud platforms to significantly reduce low resource utilization.
We introduce the \implvm{} to address the challenges of oversubscription in virtualized environments. 
By considering all resources, \pname{} provides a holistic solution that safely increases resource oversubscription, enabling cloud platforms to host up to 26\% more VMs.

%% file: acks.tex
\section*{Acknowledgements}
We thank the anonymous reviewers and our shepherd, Michael Swift, for their valuable feedback and constructive suggestions that helped improve this paper. 
We thank Arup Roy, Patrick Payne, Milos Kralj, and the entire Core OS team at Microsoft Azure for their help. 
Benjamin Reidys and Jian Huang were partially supported by NSF grant CCF-1919044 and NSF CAREER Award CNS-2144796.